\documentclass[12pt, onecolumn]{IEEEtran}

\usepackage{amsmath}
\usepackage{amssymb}
\usepackage{cite}
\usepackage{epsfig}
\usepackage{flushend}
\usepackage{multirow}

\newtheorem{thm}{Theorem}
\newtheorem{lem}[thm]{Lemma}
\newtheorem{col}{Corollary}
\newtheorem{rem}[col]{Remark}

\newcommand{\kvec}{{\rm vec}}
\newcommand{\tr}{{\rm trace}}

\newsavebox{\RR}

\begin{document}
\title{A Near Maximum Likelihood Decoding Algorithm for MIMO
Systems Based on Semi-Definite Programming}
\author{Amin Mobasher$\dag$, Mahmoud Taherzadeh$\dag$, Renata Sotirov$\ddag$, and Amir K.
Khandani$\dag$
\thanks{$\dag$ Coding \& Signal Transmission Laboratory
(www.cst.uwaterloo.ca), Dept. of Elec. and Comp. Eng., University of
Waterloo, Waterloo, Ontario, Canada, N2L 3G1, E-mail: \{amin,
taherzad, khandani\}@cst.uwaterloo.ca}
\thanks{$\ddag$ Department of Econometrics and Operations Research, Tilburg
University, Warandelaan 2, Tilburg, Netherlands, Email:
r.sotirov@uvt.nl}
\thanks{This work is financially supported by Communications and Information
Technology Ontario (CITO), Nortel Networks, and Natural Sciences and
Engineering Research Council of Canada (NSERC).}
\thanks{This work is partly presented in CISS 2005 \cite{MoTaSoKh05C1}, CWIT 2005
\cite{MoTaSoKh05C2}, and ISIT 2005 \cite{MoTaSoKh05C3}}}
\markboth{Submitted to IEEE TRANS. ON Info. Theory}{Submitted to
IEEE TRANS. ON Info. Theory} \maketitle

\begin{abstract}
In Multi-Input Multi-Output (MIMO) systems, Maximum-Likelihood (ML)
decoding is equivalent to finding the closest lattice point in an $
N $-dimensional complex space. In general, this problem is known to
be NP hard. In this paper, we propose a quasi-maximum likelihood
algorithm based on Semi-Definite Programming (SDP). We introduce
several SDP relaxation models for MIMO systems, with increasing
complexity. We use interior-point methods for solving the models and
obtain a near-ML performance with polynomial computational
complexity. Lattice basis reduction is applied to further reduce the
computational complexity of solving these models. The proposed
relaxation models are also used for soft output decoding in MIMO
systems.
\end{abstract}


\section{Introduction}

Recently, there has been a considerable interest in Multi-Input
Multi-Output (MIMO) antenna systems due to achieving a very high
capacity as compared to single-antenna systems \cite{Tel99, Fos96}.
In MIMO systems, a vector is transmitted by the transmit antennas.
In the receiver, a corrupted version of this vector affected by the
channel noise and fading is received. Decoding concerns the
operation of recovering the transmitted vector from the received
signal. This problem is usually expressed in terms of ``lattice
decoding" which is known to be NP-hard.

To overcome the complexity issue, a variety of sub-optimum
polynomial time algorithms are suggested in the literature for
lattice decoding. However, unfortunately, these algorithms usually
result in a noticeable degradation in the performance. Examples of
such polynomial time algorithms include: Zero Forcing Detector (ZFD)
\cite{Sch79,KoImHa83}, Minimum Mean Squared Error Detector (MMSED)
\cite{HoMaVe83, XiShRu90}, Decision Feedback Detector (DFD) and
Vertical Bell Laboratories Layered Space-Time Nulling and
Cancellation Detector (VBLAST Detector) \cite{GoFoVaWo99,
DeMuCoMuSiLo00}.

Lattice basis reduction has been applied as a pre-processing step in
sub-optimum decoding algorithms to reduce the complexity and achieve
a better performance. Minkowski reduction \cite{Hel85},
Korkin-Zolotarev reduction \cite{KorZol73} and LLL reduction
\cite{LLL82} have been successfully used for this purpose in
\cite{AgVaZe02, Bab86, BanKha95, BanKha98, Mow, WinFis03}.

In the last decade, Sphere Decoder (SD)\footnote{This technique is
introduced in the mathematical literature several years ago
\cite{SchEuc94, FinPoh85}.} is introduced as a Maximum Likelihood
(ML) decoding method for MIMO systems with near-optimal performance
\cite{DaChBe00le}. In the SD method, the lattice points inside a
hyper-sphere are generated and the closest lattice point to the
received signal is determined. The \textit{average complexity} of
this algorithm is shown to be polynomial time (almost cubic) over
certain ranges of rate, Signal to Noise Ratio (SNR) and dimension,
while the worst case complexity is still exponential \cite{AgVaZe02,
HasVik031, RekBel02}. However, recently, it has been shown that it
is a misconception that the expected number of operations in SD
asymptotically grows as a polynomial function of the problem size
\cite{JalOtt05} (reference \cite{JalOtt05} derives an exponential
lower bound on the average complexity of SD).

In \cite{StLuWo03}, a quasi-maximum likelihood method for lattice
decoding is introduced. Each signal constellation is expressed by
its binary representation and the decoding is transformed into a
quadratic minimization problem \cite{StLuWo03}. Then, the resulting
problem is solved using a relaxation for rank-one matrices in
Semi-Definite Programming (SDP) context. It is shown that this
method has a near optimum performance and a polynomial time worst
case complexity. However, the method proposed in~\cite{StLuWo03} is
limited to scenarios that the constellation points are expressed as
a linear combination of bit labels. A typical example is the case of
natural labeling in conjunction with PSK constellation. This
restriction is removed in this work. Therefore, we are able to
handle any form of constellation for an arbitrary labeling of
points, for example PAM constellation with Gray labeling\footnote{It
is shown that Gray labeling, among all possible constellation
labeling methods, offers the lowest possible average probability of
bit errors \cite{AgLaStOt04}.}. Another quasi-maximum likelihood
decoding method is introduced in \cite{LuLuKi03} for larger PSK
constellations with near ML performance and low complexity.

Recently, we became aware of another quasi-maximum likelihood
decoding method \cite{WiElSh05L} for the MIMO systems employing
16-QAM. They replace any finite constellation by a polynomial
constraint, e.g. if $x\in\lbrace a,b,c\rbrace$, then
$(x-a)(x-b)(x-c)=0$. Then, by introducing some slack variables, the
constraints are expressed in terms of quadratic polynomials.
Finally, the SDP relaxation is resulted by dropping the rank-one
constraint. The work in \cite{WiElSh05L} , in its current form, is
restricted to MIMO systems employing 16-QAM. However, it can be
generalized for larger constellations at the cost of defining more
slack variables, increasing the complexity, and significantly
decreasing the performance\footnote{Private Communication with
Authors}.

In this work, we develop an efficient approximate ML decoder for
MIMO systems based on SDP. In the proposed method, the transmitted
vector is expanded as a linear combination (with zero-one
coefficients) of all the possible constellation points in each
dimension. Using this formulation, the distance minimization in
Euclidean space is expressed in terms of a binary quadratic
minimization problem. The minimization of this problem is over the
set of all binary rank-one matrices with column sums equal to one.
In order to solve this minimization problem, we present two
relaxation models (Model III and IV), providing a trade-off between
the computational complexity and the performance (both models can be
solved with polynomial-time complexity). Two additional relaxation
models (Model I and II) are presented as intermediate steps in the
derivations of Model III and IV.

{\bf Model I}: A preliminary SDP relaxation of the minimization
problem is obtained by removing the rank-one constraint in the
problem and using Lagrangian duality \cite{WoSaVa00}. This
relaxation has many redundant constraints and no strict interior
point in the feasible set (there are numerical difficulties in
computing the solution for a problem without an interior point).

{\bf Model II}: To overcome this drawback, the feasible set is
projected onto a face of the semi-definite cone. Then, based on the
identified redundant constraints, another form of the relaxation is
obtained, which can be solved using interior-point methods.

{\bf Model III}: The relaxation Model II results in a weak lower
bound. To strengthen this relaxation model, the structure of the
feasible set is investigated. An interesting property of the
feasible set imposes a zero pattern for the solution. Adding this
pattern as an extra constraint to the previous relaxation model
results in a stronger model.

{\bf Model IV}: Finally, the strongest relaxation model in this work
is introduced by adding some additional non-negativity constraints.
The number of non-negativity constraints can be adjusted to provide
a trade-off between the performance and complexity of the resulting
method.

Simulation results show that the performance of the last model is
near optimal for M-ary QAM or PSK constellation (with an arbitrary
binary labeling, say Gray labeling). Therefore, the decoding
algorithm built on this model has a near-ML performance with
polynomial computational complexity.

The proposed models result in a solution that is not necessarily a
binary rank-one matrix. This solution is changed to a binary
rank-one matrix through a randomization algorithm. We modify the
conventional randomization algorithms to adopt to our problem. Also,
a new randomization procedure is introduced which finds the optimal
binary rank-one solution in a smaller number of iterations than the
conventional ones. Finally, we discuss using a lattice basis
reduction method to further reduce the computational complexity of
the proposed relaxation models. The extension of the decoding
technique for soft output decoding is also investigated.

Following notations are used in the sequel. The space of $ K \times
N $ (resp.~$ N \times N $) real matrices is denoted by ${\mathcal
M}_{K \times N}$ (resp.~${\mathcal M}_{N}$), and the space of $ N
\times N $ symmetric matrices is denoted by $ {\mathcal S}_N $. The
indexing of the matrix elements start from one, unless in cases that
an extra row (or column) is inserted in the first row (or column) of
a matrix in which case indexing starts from zero. For a $K\times N$
matrix $ \mathbf{X} \in {\mathcal M}_{K \times N} $ the $(i,j)$th
element is represented by $x_{ij}$, where $1\leq i \leq K,\; 1\leq j
\leq N$, i.e. ${\bf X}=[x_{ij}]$. We use ${\tr}(\mathbf{A})$ to
denote the trace of a square matrix ${\bf A}$. The space of
symmetric matrices is considered with the trace inner product
$\langle \mathbf{A},\mathbf{B} \rangle = {\tr}(\mathbf{AB})$. For $
\mathbf{A}, \mathbf{B} \in {\mathcal S}_{N}$, $ \mathbf{A} \succeq
0$ (resp.~$ \mathbf{A} \succ 0$) denotes positive semi-definiteness
(resp. positive definiteness), and $ \mathbf{A} \succeq \mathbf{B}$
denotes  $ \mathbf{A}-\mathbf{B} \succeq 0$. For two matrices $
\mathbf{A}, \mathbf{B} \in {\mathcal M}_{N}$, $ \mathbf{A} \geq
\mathbf{B} $, ($ \mathbf{A} > \mathbf{B} $) means $a_{ij}\geq
b_{ij}$, $(a_{ij}
> b_{ij})$ for all $i,j$. The Kronecker product of two matrices $
\mathbf{A} $ and $ \mathbf{B} $ is denoted by $ \mathbf{A} \otimes
\mathbf{B} $ (for definition see \cite{Gra81}).

For $ \mathbf{X} \in {\mathcal M}_{K\times N}$, ${\rm
vec}(\mathbf{X})$ denotes the vector in $\mathbb{R}^{KN}$ (real
$KN$-dimensional space) that is formed from the columns of the
matrix $ \mathbf{X} $. The following identity relates the Kronecker
product with ${\rm vec}(\cdot)$ operator, see e.\,g. \cite{Gra81},
\begin{equation}\label{veckron}
{\rm vec}(\mathbf{ACB}) = (\mathbf{B}^{T}\otimes \mathbf{A}){\rm
vec}(\mathbf{C}).
\end{equation}
For $\mathbf{X}\in {\mathcal M}_{N}$, ${\rm diag}(\mathbf{X})$ is a
vector of the diagonal elements of $\mathbf{X}$. We use
$\mathbf{e}_N \in \mathbb{R}^N$ (resp. $ \mathbf{0}_N  \in
\mathbb{R}^N $) to denote the $ N \times 1 $ vector of all ones
(resp. all zeros), $ \mathbf{E}_{K\times N} \in {\mathcal M}_{K
\times N}$ to denote the matrix of all ones, and $ \mathbf{I}_N $ to
denote the $N\times N$ Identity matrix. For $ \mathbf{X} \in
{\mathcal M}_{K \times N}$, the notation $ \mathbf{X}(1:i,1:j) $,
$i<k$ and $j<n$ denotes the sub-matrix of $ \mathbf{X}$ containing
the first $i$ rows and the first $j$ columns.

The rest of the paper is organized as follows. The problem
formulation is introduced in Section II. Section III is devoted to
the semi-definite solution of this problem. In Section IV,
randomization procedures for finding rank-one solutions are
presented. Section V introduces a method based on lattice basis
reduction to reduce the computational complexity of the proposed
relaxation models. In Section VI, the soft decoding methods based on
the proposed models are investigated. Simulation results are
presented in Section VII. Finally, Section VIII concludes the paper.

\section{Problem Formulation}

A MIMO system with $\tilde N$ transmit antennas and $\tilde M$
receive antennas is modelled as
\begin{equation}\label{eq:compchan}
\tilde{\mathbf{y}}=\sqrt{\dfrac{S\!N\!R}{\tilde{M}
\tilde{E}_{s_{av}}}} \tilde{\mathbf{H}} \tilde{\mathbf{x}} +
\tilde{\mathbf{n}} ,
\end{equation}
where $ \tilde{\mathbf{H}}=\left[ \tilde{h}_{ij} \right] $ is the $
\tilde{M} \times \tilde{N} $ channel matrix composed of independent,
identically distributed complex Gaussian random elements with zero
mean and unit variance, $ \tilde{ \mathbf{n}} $ is an $ \tilde{M}
\times 1 $ complex additive white Gaussian noise vector with zero
mean and unit variance, and $ \tilde{\mathbf{x}} $ is an $ \tilde{N}
\times 1 $ data vector whose components are selected from a complex
set $ \tilde{\mathcal{S}} = \left\lbrace \tilde{s}_1, \tilde{s}_2,
\cdots, \tilde{s}_K \right\rbrace $ with an average energy of $
\tilde{E}_{s_{av}} $. The parameter $S\!N\!R$ in (\ref{eq:compchan})
is the SNR per receive antenna.

Noting $ \tilde{x}_i \in \tilde{\mathcal{S}} $, for $
i=1,\cdots,\tilde{N} $, we have
\begin{equation}\label{eq:ideac}
\tilde{x}_i=u_i(1)\tilde{s}_1 + u_i(2)\tilde{s}_2 + \cdots +
u_i(K)\tilde{s}_K,
\end{equation}
where
\begin{equation}\label{eq:idcond}
u_i(j) \in \left\lbrace 0,1 \right\rbrace \;\; {\rm and } \;\;
\sum_{j=1}^K u_i(j)=1, \;\; \forall\, i=1,\cdots,\tilde{N}.
\end{equation}
Let
\begin{math}
{\bf u} = \left[\hspace{-2pt}
\begin{array}{ccccccc}
u_1(1) & \hspace{-2pt} \cdots \hspace{-2pt} & u_1(K) & \hspace{-2pt}
\cdots \hspace{-2pt} & u_{N}(1) & \hspace{-2pt} \cdots \hspace{-2pt}
& u_{N}(K)
\end{array}\hspace{-2pt}
\right]^T
\end{math} and $N=\tilde{N}$.
Using the equations in (\ref{eq:ideac}) and (\ref{eq:idcond}), the
transmitted vector is expressed as
\begin{equation}
\tilde{\mathbf{x}}=\tilde{\mathbf{S}}\mathbf{u},
\end{equation}
where $ \tilde{\mathbf{S}} = {\bf I}_N \otimes [\tilde
s_1,\cdots,\tilde s_K] $ is an $ N \times NK $ matrix of
coefficients, and $ \mathbf{u} $ is an $ NK \times 1 $ binary vector
such that $ \mathbf{A}\mathbf{u}=\mathbf{e}_N $, where $ \mathbf{A}
= {\bf I}_N \otimes {\bf e}_K^T $. This constraint states that among
each $K$ components of the binary vector ${\bf u}$, i.e. $ u_i(1),
\cdots, u_i(K) $, there is only one element equal to ``1" and the
rest are zero.

To avoid using complex matrices, the system model
(\ref{eq:compchan}) is represented by real matrices in
(\ref{eq:realchan}).
\begin{eqnarray}\label{eq:realchan}
\nonumber \left[\hspace{-5pt}
\begin{tabular}{c}
$ \mathfrak{R}\left( \tilde{\mathbf{y}} \right) $ \\
$ \mathfrak{I}\left( \tilde{\mathbf{y}} \right) $ \\
\end{tabular}\hspace{-5pt}
\right] & = & \sqrt{\dfrac{SNR}{\tilde{M} \tilde{E}_{s_{av}}}}
\left[\hspace{-5pt}
\begin{tabular}{cc}
$ \mathfrak{R}\left( \tilde{\mathbf{H}} \right) $ & $
\mathfrak{I}\left( \tilde{\mathbf{H}} \right) $ \\
$ -\mathfrak{I}\left( \tilde{\mathbf{H}} \right) $ & $
\mathfrak{R}\left( \tilde{\mathbf{H}} \right) $ \\
\end{tabular} \hspace{-5pt}
\right] \left[ \hspace{-5pt}
\begin{tabular}{c}
$ \mathfrak{R}\left( \tilde{\mathbf{x}} \right) $ \\
$ \mathfrak{I}\left( \tilde{\mathbf{x}} \right) $ \\
\end{tabular} \hspace{-5pt}
\right]\\
\nonumber & + & \left[ \hspace{-5pt}
\begin{tabular}{c}
$ \mathfrak{R}\left( \tilde{\mathbf{n}} \right) $ \\
$ \mathfrak{I}\left( \tilde{\mathbf{n}} \right) $ \\
\end{tabular} \hspace{-5pt}
\right]\\
\Rightarrow \mathbf{y} & = & \mathbf{H} \mathbf{x} + \mathbf{n},
\end{eqnarray}
where $ \mathfrak{R}(.) $ and $ \mathfrak{I}(.) $ denote the real
and imaginary parts of a matrix, respectively, $ \mathbf{y} $ is the
\emph{received vector}, and $ \mathbf{x} $ is the \emph{input
vector}.\\
Let $ \mathbf{S} $ denotes the real matrix
\begin{math}
\left[\hspace{-5pt}
\begin{tabular}{c}
$ \mathfrak{R} \left( \tilde{\mathbf{S}} \right) $\\
$ \mathfrak{I} \left( \tilde{\mathbf{S}} \right) $
\end{tabular}\hspace{-5pt}
\right];
\end{math}
therefore,
\begin{equation}\label{eq:binchan}
\mathbf{y} = \mathbf{H} \mathbf{S} \mathbf{u} + \mathbf{n}
\end{equation}
expresses the MIMO system model by real matrices and the input
binary data vector, ${\bf u}$.

Consider the case that different components of ${\bf \tilde{x}}$ in
(\ref{eq:compchan}), corresponding to the two-dimensional
sub-constellations, are equal to the cartesian product of their
underlying one-dimensional sub-constellations, e.g. QAM signalling.
In this case, the components of ${\bf x}$ in (\ref{eq:realchan})
belong to the set $ \mathcal{S} = \left\lbrace s_1, \cdots, s_K
\right\rbrace $ with real elements, i.e.
\begin{equation}\label{eq:idear}
x_i=u_i(1)s_1 + u_i(2)s_2 + \cdots + u_i(K)s_K,
\end{equation}
where only one of the $ u_i(j) $ is $ 1 $ and the rest are zero.

Let $ \mathbf{u} = \left[ u_1(1) \cdots u_1(K) \cdots u_N(1) \cdots
u_N(K) \right]^T $, $N=2\tilde{N}$, and $ \mathbf{S} = {\bf I}_N
\otimes [s_1,\cdots,s_K] $. Then, the equation for the components of
${\bf x}$ in (\ref{eq:idear}) reduces to $
\mathbf{x}=\mathbf{S}\mathbf{u} $ and the relationship for the MIMO
system model is given in (\ref{eq:binchan}).

At the receiver, the Maximum-Likelihood (ML) decoding rule is given
by
\begin{equation}\label{eq:harddec}
\hat{\mathbf{x}}=\arg \min_{x_i\in \mathcal{S}} \| \hat{\mathbf{y}}
- \mathbf{H} \mathbf{x} \|^2,
\end{equation}
where $ \hat{\mathbf{x}} $ is the most likely input vector and $
\hat{\mathbf{y}} $ is the received vector. Noting ${\bf x}={\bf
Su}$, this problem is equivalent to
\begin{eqnarray}
\nonumber &&\min_{\mathbf{A}\mathbf{u}=\mathbf{e}_N} \|
\hat{\mathbf{y}}-\mathbf{H}\mathbf{S}\mathbf{u} \|^2 \equiv\\
&&\min_{\mathbf{A}\mathbf{u}=\mathbf{e}_N} \mathbf{u}^T \mathbf{S}^T
\mathbf{H}^T \mathbf{H} \mathbf{S} \mathbf{u} - 2\hat{\mathbf{y}}^T
\mathbf{H} \mathbf{S} \mathbf{u},
\end{eqnarray}
where ${\bf u}$ is a binary vector.

Let $ \mathbf{Q} = \mathbf{S}^T \mathbf{H}^T \mathbf{H} \mathbf{S} $
and $ \mathbf{c} = -\mathbf{S}^T \mathbf{H}^T \hat{\mathbf{y}} $.
Therefore, this problem is formulated as
\begin{align}\label{eq:semiprob}
\nonumber \min \;\; &\mathbf{u}^T \mathbf{Q}
\mathbf{u} + 2 \mathbf{c}^T \mathbf{u} \\
\nonumber
s.t. \;\; &\mathbf{A}\mathbf{u}=\mathbf{e}_N \\
&{\bf u} \in \left\lbrace 0,1 \right\rbrace^n ,
\end{align}
where $ n = NK $. The formulation (\ref{eq:semiprob}) is a quadratic
minimization problem with binary variables \cite{WoSaVa00}. Recent
studies on solving binary quadratic minimization problems such as
Graph Partitioning \cite{WolZha99} and Quadratic Assignment Problem
\cite{ZhKaReWo96, SotRen03}, show that semi-definite programming is
a very promising approach to provide tight relaxations for such
problems. In the following, we derive several SDP relaxation models
for the minimization problem in (\ref{eq:semiprob}). Appendix
\ref{app:lag} provides the mathematical framework for these models
using the Lagrangian Duality \cite{WoSaVa00}.

\section{Semi-Definite Programming Solution}\label{sec:semi}

Consider the minimization problem in (\ref{eq:semiprob}). Since
${\bf u}$ is a binary vector, the objective function is expressed as
\begin{eqnarray}
\nonumber {\bf u}^T  \mathbf{Q} {\bf u} + 2\mathbf{c}^T{\bf u} &=&
\tr \left( \left [
\begin{array}{cc}
1 & {\bf u}^T
\end{array}
\right ] {\bf {\mathcal L}_Q} \left [
\begin{array}{c}
1 \\ {\bf u}
\end{array}
\right ] \right )\\
\nonumber&=& \tr \left( {\bf {\mathcal L}_Q} \left [
\begin{array}{c}
1 \\ {\bf u}
\end{array}
\right ]\left [
\begin{array}{cc}
1 & {\bf u}^T
\end{array}
\right ] \right )\\ &=& \tr \left( {\bf {\mathcal L}_Q} \left [
\begin{array}{c|c}
1 & {\bf u}^T \\ \hline {\bf u} & {\bf u} {\bf u}^T
\end{array}
\right ]\right),
\end{eqnarray}
where
\begin{math}
{\bf {\mathcal L}_Q} := \left [
\begin{array}{c|c}
0 & \mathbf{c}^T \\ \hline \mathbf{c} & \mathbf{Q}
\end{array}
\right ].
\end{math}

Let ${\mathcal E}_{K \times N}$ denote the set of all binary
matrices in ${\mathcal M_{K\times N}}$ with column sums equal to
one, i.e.
\begin{eqnarray}
{\mathcal E}_{K \times N} \hspace{-2pt} = \hspace{-2pt} \left\lbrace
{\bf X} \hspace{-2pt} \in \hspace{-2pt} {\mathcal M_{K\times N}}:
{\bf e}^T_K {\bf X} = {\bf e}^T_N, x_{ij} \in \{0,1\}, \forall i,j
\right\rbrace.
\end{eqnarray}
Since the constraints $ {\bf A}{\bf u} = \mathbf{e}_N $, $u_i \in \{
0,1 \}^{NK}$ in (\ref{eq:semiprob}) and ${\bf u} = \kvec ({\bf U}),
{\bf U} \in \mathcal{E}_{K\times N} $ are equivalent, the
minimization problem (\ref{eq:semiprob}) can be written as
\begin{equation}\label{eq:semisum}
\begin{array}{ll}
\min & \tr ~{{\mathcal L}_Q} \left [
\begin{array}{c|c}
1 & {\bf u}^T \\ \hline {\bf u} & {\bf u} {\bf u}^T
\end{array}
\right ] \\ [2.5ex] {\rm s.t.} & {\bf u} = \kvec({\bf U}), ~{\bf
U}\in {\mathcal E}_{K\times N}.
\end{array}
\end{equation}

To derive the first semi-definite relaxation model, a direct
approach based on the well known lifting process \cite{BaCeCo93} is
selected. In accordance to (\ref{eq:semisum}), for any ${\bf U}\in
{\mathcal E}_{K\times N}$, ${\bf u} = \kvec({\bf U})$, the feasible
points of (\ref{eq:semisum}) are expressed by
\begin{equation}\label{eq:optY}
{\bf Y}_{\bf u} = \left [
\begin{array}{c}
1 \\ {\bf u}
\end{array}
\right ]\left [
\begin{array}{cc}
1 & {\bf u}^T
\end{array}
\right ] = \left [
\begin{array}{c|c}
1 & {\bf u}^T \\ \hline {\bf u} & {\bf u} {\bf u}^T
\end{array}
\right ].
\end{equation}
The matrix ${\bf Y}_{\bf u}$ is a rank-one and positive
semi-definite matrix. Also, we have
\[
\begin{array}{l}
{\rm diag}({\bf Y}_{\bf u})={\bf Y}^T_{{\bf u}_{0,:}}={\bf Y}_{{\bf
u}_{:,0}},
\end{array}
\]
where ${\bf Y}_{{\bf u}_{0,:}}$ (resp. ${\bf Y}_{{\bf u}_{:,0}}$)
denotes the first row (resp. the first column)\footnote{Matrix ${\bf
Y}_{\bf u}$ is indexed from zero.} of ${\bf Y_{\bf u}}$ (Note that
${\bf u}$ is a binary vector, and consequently, ${\rm diag}({\bf
uu}^T)={\bf u}$).

In order to obtain a tractable SDP relaxation of (\ref{eq:semisum}),
we remove the rank-one restriction from the feasible set. In fact,
the feasible set is approximated by another larger set $ \mathcal F
$, defined as
\begin{equation}
{\mathcal F} := {\rm conv}\left \{ {\bf Y}_{\bf u}: {\bf u} =
\kvec({\bf U}), ~{\bf U}\in {\mathcal E}_{K\times N} \right \},
\end{equation}
where ${\rm conv(.)}$ denotes the convex hull of a set. This results
in our first relaxation model ({\bf Model I}) for the original
problem given in (\ref{eq:semiprob}):
\begin{equation}\label{eq:semiY}
\begin{array}{ll}
\min & ~\tr {\bf \mathcal{L}_Q Y}\\
{\rm s.t.} & {\bf Y} \in {\mathcal F}
\end{array}
\end{equation}

It is clear that the matrices $$ {\bf Y}_{\bf u} {\rm ~for~} {\bf u}
= \kvec({\bf U}), ~{\bf U}\in {\mathcal E}_{K\times N} $$ are the
feasible points of ${\mathcal F}$. Moreover, since these points are
rank-one matrices, they are contained in the set of extreme points
of ${\mathcal F}$, see e.g. \cite{Pat94}. In other words, if the
matrix $ \mathbf{Y} $ is restricted to be rank-one in
(\ref{eq:semiY}), i.e.
\begin{math}
\mathbf{Y}=\left[
\begin{tabular}{c}
$ 1 $\\
$ \mathbf{u} $\\
\end{tabular}
\right] \left[1 \; \mathbf{u}^T \right],
\end{math}
for some $ \mathbf{u} \in \mathbb{R}^n $, then the optimal solution
of (\ref{eq:semiY}) provides the optimal solution of
(\ref{eq:semiprob}).

%
The SDP relaxation problem (\ref{eq:semiY}) is not solvable in
polynomial time and $\mathcal F$ has no interior points. Therefore
our goal is to approximate the set $\mathcal F$ by a larger set
containing $\mathcal F$. In the following, we show that $\mathcal F$
actually lies in a smaller dimensional subspace. We will further see
that relative to this subspace, $\mathcal F$ will have interior
points.

\subsection{Geometry of the Relaxation}

In order to approximate the feasible set $\mathcal{F}$ for solving
the problem, we elaborate more on the geometrical structure of this
set. First, we prove the following lemma on the representation of
matrices having sum of the elements in each column equal to one.
\begin{lem} \label{new} Let
\begin{equation} \label{def:V}
{\bf V}_{K \times (K-1)} = \left [
\begin{array}{c}
{\bf I}_{K-1} \\ \hline -{\bf e}_{K-1}^{T}
\end{array} \right ] \in {\mathcal M}_{K \times (K-1)}
\end{equation}
and
\begin{equation} \label{def:F}
{\bf F}_{K\times N} := \frac{1}{K} \left[ {\bf E}_{K\times N} - {\bf
V}_{K \times (K-1)} {\bf E}_{(K-1)\times N}\right].
\end{equation}
A matrix ${\bf X} \in {\mathcal M}_{K \times N}$ with the property
that the summation of its elements in each column is equal to one,
i.e. $ {\bf e}^T_K {\bf X} = {\bf e}^T_N$, can be written as
\begin{equation}
{\bf X} = {\bf F}_{K\times N} + {\bf V}_{K \times (K-1)} {\bf Z},
\end{equation}
where $ {\bf Z} = {\bf X}(1:(K-1),1:N)$.
\end{lem}

\noindent
\begin{proof}
see Appendix \ref{app:proof}.
\end{proof}
\begin{col}
$\forall {\bf X} \in {\mathcal E}_{K\times N}$, $ \exists {\bf Z}\in
{\mathcal M}_{(K-1)\times N}$, $z_{ij}\in \{0,1\}$ s.t. ${\bf
X}={\bf F}_{K\times N}+{\bf V}_{K}{\bf Z}$, where ${\bf Z}={\bf X}
(1:(K-1),1:N)$.
\end{col}
Using Lemma \ref{new}, the following theorem can be proved which
provides the structure of the elements in the set $\mathcal{F}$.
\begin{thm} \label{lem:SetP}
Let
\begin{equation} \label{VTT:def}
\hat {\bf V} = \left [
\begin{array}{c|c} 1 & {\bf 0}_{N(K-1)}^T
\\ \hline
 \frac{1}{K} ( {\bf e}_{NK} - ({\bf I}_N \otimes {\bf V}_{K \times (K-1)}) {\bf e}_{(K-1)N})
 &
 {\bf I}_N \otimes {\bf V}_{K \times (K-1)} \end{array} \right ],
\end{equation}
where $\hat {\bf V} \in {\mathcal M}_{ (NK+1) \times ((K-1)N+1)}$.
For any ${\bf Y} \in {\mathcal F}$, there exists a symmetric matrix
${\bf R}$ of order $N(K-1)+1$, indexed from 0 to $N(K-1)$, such that
\begin{equation}
 {\bf Y} = \hat {\bf V} {\bf R} \hat {\bf V}^{T}, ~{\bf R} \succeq 0,~ \mbox{ and } ~r_{00}=1, ~r_{ii}=r_{0i}, ~\forall i.
\end{equation}
Also, if  ${\bf Y}$ is an extreme point of ${\mathcal F}$, then
$r_{ij}\in \{ 0,1\}$, otherwise $r_{ij} \in [0,1]$ for $i,j \in
\{0,\ldots, N(K-1) \}$.
\end{thm}
\begin{proof}
see Appendix \ref{app:proof}.
\end{proof}
Using Theorem \ref{lem:SetP}, we can show that the set
$\mathcal{F}_r$ contains ${\mathcal F}$:
\begin{eqnarray}
\nonumber\mathcal{F}_r = \left \{ {\bf Y} \in {\mathcal S}_{NK +1}:
\exists \, {\bf R} \in {\mathcal  S}_{(K-1)N+1}, ~{\bf R} \succeq
{\bf 0},\right. \\
\left.  ~R_{00}=1, {\bf Y}=\hat {\bf VR} \hat {\bf V}^T, ~{\rm
diag}({\bf Y})={\bf Y}_{0,:} \right \}.
\end{eqnarray}
Therefore, the feasible set in (\ref{eq:semiY}) is approximated by
$\mathcal{F}_r$. This results in our second relaxation model ({\bf
Model II}) of the original problem given in (\ref{eq:semiprob}):
\begin{align}\label{eq:SemiR}
\nonumber \min \;\; &\textmd{trace }
(\mathbf{\hat{V}}^T\mathcal{L}_{\mathbf{Q}} \mathbf {\hat{V}})
\mathbf{R}\\
\nonumber \textmd{s.t. }\; &\textmd{diag} (\mathbf{\hat{V}R
\hat{V}}^T) = (1, (\mathbf{\hat{V}R\hat{V}}^T)_{0,1:n})^T\\ &
\mathbf{R} \succeq 0.
\end{align}

Note that the matrices ${\bf Y}_{\bf u}$ are contained in the set of
extreme points of $\mathcal{F}$. We need only consider faces of
$\mathcal{F}$ which contain all of these extreme points. Therefore,
we are only looking for the \textit{minimal face}, which is the
intersection of all these faces. We will show later that the SDP
relaxation (\ref{eq:SemiR}) is the projection of the SDP relaxation
(\ref{eq:semiY}) onto the minimal face of $\mathcal{F}$.

Solving the relaxation model in (\ref{eq:semiY}) over $\mathcal{F}$
results in the optimal solution of the original problem in
(\ref{eq:semisum}), but this problem is NP-hard. Solving the
relaxation model in (\ref{eq:SemiR}) over $\mathcal{F}_r$ results in
a weaker bound for the optimal solution. In order to improve this
bound, the relaxation is strengthen by adding an interesting
property of the matrix ${\bf Y}_{\bf u}$. This results in the next
relaxation model.

\subsection{Tightening the Relaxation by Gangster Operator}

The feasible set of the minimization problem (\ref{eq:SemiR}) is
convex. It contains the set of matrices of the form ${\bf Y}_{\bf u}
$ corresponding to different vectors $ {\bf u} $. However, the SDP
relaxations may contain many points that are not in the affine hull
of these ${\bf Y}_{\bf u} $. In the following, we extract a
condition which is implicit in the matrix ${\bf Y}_{\bf u} $ and
explicitly add it to the relaxation model (\ref{eq:SemiR}).
Subsequently, some redundant constraint are removed and this results
in an improved relaxation (relaxation \emph{Model III}).

\begin{thm}\label{thm:bary}
Let $ \mathcal{U} $ denote the set of all binary vectors $ {\bf u} =
\kvec({\bf U}), ~{\bf U}\in {\mathcal E}_{K\times N} $. Define the
\emph{barycenter point}, $ \mathbf{\hat{Y}} $, as the arithmetic
mean of all the feasible points in the minimization problem
(\ref{eq:semisum}); therefore,
\begin{equation}
\mathbf{\hat{Y}} = \dfrac{1}{K^N} \sum_{{\bf u} \in \mathcal{U}}
{\bf Y}_{\bf u} = \dfrac{1}{K^N} \sum_{{\bf u} \in \mathcal{U}}
\left[
\begin{tabular}{c|c}
$ 1 $ & $ \mathbf{u}^T $\\ \hline $ \mathbf{u} $ & $
\mathbf{u}\mathbf{u}^T $
\end{tabular} \right].
\end{equation}
Then:
\begin{enumerate}
\item [i)]$ \mathbf{\hat{Y}} $ has (a) the value of $1$ as its $ (0,0)
$ element, (b) $N$ blocks of dimension $ K \times K $ on its
diagonal which are diagonal matrices with elements $ 1/K $, and (c)
the first row and first column equal to the vector of its diagonal
elements. The rest of the matrix is composed of $ K \times K $
blocks with all elements equal to $ 1/K^2 $:
\begin{eqnarray}
\nonumber \mathbf{\hat{Y}} &=& \left[ \hspace{-5pt}
\begin{array}{c|c}
1 & \frac{1}{K}\mathbf{e}^T_n \\
\hline \frac{1}{K}\mathbf{e}_n & \hspace{-5pt}
\begin{array}{cccc}
\frac{1}{K}\mathbf{I}_K & \frac{1}{K^2} \mathbf{E}_K & \cdots & \frac{1}{K^2} \mathbf{E}_K \\
\vdots & \vdots & \ddots & \vdots \\
\vdots & \vdots & \ddots & \vdots \\
\frac{1}{K^2} \mathbf{E}_K & \cdots & \frac{1}{K^2} \mathbf{E}_K & \frac{1}{K}\mathbf{I}_K \\
\end{array}
\hspace{-5pt}
\end{array}\hspace{-5pt}\right]\\[1.5ex]
\nonumber&=&\left[
\begin{tabular}{c}
$ 1 $\\
$ \frac{1}{K}\mathbf{e}_n $
\end{tabular}\right]\left[
\begin{tabular}{cc}
$ 1 $ & $ \frac{1}{K}\mathbf{e}^T_n $
\end{tabular} \right]\\[1.5ex]
&+& \left[
\begin{array}{c|c}
0 & {\bf 0}_n^T\\
\hline {\bf 0}_n & \frac{1}{K^2} \mathbf{I}_N \otimes \left(
K\mathbf{I}_K - \mathbf{E}_K \right)
\end{array}\right];
\end{eqnarray}

\item [ii)]
\begin{math}
\textmd{rank}(\mathbf{\hat{Y}}) = N(K-1)+1;
\end{math}

\item [iii)] The $NK +1$ eigenvalues of $ \mathbf{\hat{Y}} $ are given in
the vector

\begin{center}
\begin{math}
\left( \dfrac{K+N}{K}, \dfrac{1}{K}\mathbf{e}^T_{_{N(K-1)}},{\bf
0}_N^T \right)^T;
\end{math}
\end{center}

\item [iv)]
The null space of $\mathbf{\hat{Y}}$ can be expressed by
\begin{math}
\mathcal{N}(\mathbf{\hat{Y}}) = \left\lbrace u:u\in
\mathcal{R}(\mathbf{T}^T)\right\rbrace ,
\end{math}
where the constraint matrix $ \mathbf{T} $ is the following $ N
\times (NK+1) $ matrix
\begin{displaymath}
\mathbf{T}=\left[
\begin{tabular}{c|c}
$ -\mathbf{e}_N $ & $ \mathbf{A}$
\end{tabular} \right];
\end{displaymath}

\item [v)] the range of $ \mathbf{\hat{Y}} $ can be expressed by the columns of
the $ (NK+1) \times (N(K-1)+1) $ matrix $\mathbf{\hat{V}}$.
Furthermore, $ \mathbf{T\hat{V}} = 0 $.
\end{enumerate}
\end{thm}
\begin{proof}
see Appendix \ref{app:proof}.
\end{proof}
\begin{rem}
The faces of the positive semi-definite cone are characterized by
the null space of the points in their relative interior. The minimal
face of the SDP problem contains matrices ${\bf Y}_{\bf u}$ and can
be expressed as $ \hat{\bf V}\mathcal{S}_{N(K-1)+1}\hat{\bf V}^T$.
Thus, the SDP relaxation (\ref{eq:SemiR}) is a projected relaxation
onto the minimal face of the feasible set $\mathcal{F}$.
\end{rem}

Theorem \ref{thm:bary} suggests a zero pattern for the elements of
$\mathcal F$. We use a \textit{Gangster Operator} \cite{ZhKaReWo96}
to represent these constraints more efficiently. Let $J$ be a set of
indices, then this operator is defined as

\begin{equation}
\left( \mathcal{G}_J (\mathbf{Y}) \right)_{ij} = \left\lbrace
\begin{tabular}{ll}
$ Y_{ij} $ & $ \textmd{if } (i, j) \textmd{ or } (j, i) \in J $ \\
$ 0 $ & $ \textmd{otherwise} $.
\end{tabular}
\right.
\end{equation}
Considering the barycenter point, we have $
\mathcal{G}_J(\mathbf{\hat{Y}}) =0 $ for
\begin{align}\label{eq:gangbary}
\nonumber J = \left\lbrace (i,j): \; i=K(p-1)+q, \; j=K(p-1) + r,\right.\\
\left. q<r, q,r \in \lbrace 1,\cdots,K \rbrace , p \in \lbrace 1,
\cdots, N\rbrace \right\rbrace .
\end{align}
Since $ \mathbf{\hat{Y}} $ is a convex combination of all matrices
in $ \mathcal{U} $ with entries either $ 0 $ or $ 1 $; hence, from
(\ref{eq:gangbary}), we have $ \mathcal{G}_J(\mathbf{Y}_{\bf u}) =0
$. Also, all the points from the feasible set $ \mathcal{F} $ are
the convex combination of ${\bf Y}_{\bf u}$. Therefore,
\begin{equation}
\mathcal{G}_J(\mathbf{Y}) =0, \qquad \forall \mathbf{Y} \in
\mathcal{F}.
\end{equation}
The feasible set of the projected SDP in (\ref{eq:SemiR}) is
tightened by adding the constraints $ \mathcal{G}_J (\mathbf{Y}) = 0
$. By combining these constraints and (\ref{eq:SemiR}), we note that
there are some redundant constraints that can be removed to enhance
the relaxation model. This is expressed in the following lemma.

\begin{lem}\label{lem:cons}
Let $ \mathbf{R} $ be an arbitrary $ (N(K-1)+1) \times (N(K-1)+1) $
symmetric matrix with
\begin{equation}
\mathbf{R}=\left[
\begin{tabular}{c|ccc}
$ r_{00} $ & $ \mathbf{R}_{01} $ & $ \cdots $ & $ \mathbf{R}_{0N} $ \\
\hline
$ \mathbf{R}_{10} $ & $ \mathbf{R}_{11} $ & $ \cdots $ & $ \mathbf{R}_{1N} $ \\
$ \vdots $ & $ \ddots $ & $ \ddots $ & $ \vdots $ \\
$ \mathbf{R}_{N0} $ & $ \mathbf{R}_{N1} $ & $ \cdots $ & $
\mathbf{R}_{NN} $
\end{tabular}\right],
\end{equation}
where $ r_{00} $ is a scalar, $ \mathbf{R}_{i0} $, for $
i=1,\cdots,N $ are $ (K-1) \times 1 $ vectors and $ \mathbf{R}_{ij}
$, for $ i,j=1,\cdots,N $, are $ (K-1) \times (K-1) $ blocks of
$\mathbf{R} $. Theorem \ref{lem:SetP} states that $ \mathbf{Y} =
\mathbf{\hat{V}R}{\bf \hat{V}}^T $. We can partition $ \mathbf{Y} $
as
\begin{equation}
\mathbf{Y}=\left[
\begin{tabular}{c|ccc}
$ y_{00} $ & $ \mathbf{Y}_{01} $ & $ \cdots $ & $ \mathbf{Y}_{0N} $ \\
\hline
$ \mathbf{Y}_{10} $ & $ \mathbf{Y}_{11} $ & $ \cdots $ & $ \mathbf{Y}_{1N} $ \\
$ \vdots $ & $ \ddots $ & $ \ddots $ & $ \vdots $ \\
$ \mathbf{Y}_{N0} $ & $ \mathbf{Y}_{N1} $ & $ \cdots $ & $
\mathbf{Y}_{NN} $
\end{tabular}\right],
\end{equation}
where $ y_{00} $ is a scalar, $ \mathbf{Y}_{i0} $, for $
i=1,\cdots,N $ are $ K \times 1 $ vectors and $ \mathbf{Y}_{ij} $,
for $ i,j=1,\cdots,N $, are $ K \times K $ blocks of $ \mathbf{Y} $.
Then,
\begin{enumerate}
\item $ y_{00} = r_{00} $ and $ \mathbf{Y}_{0i}\mathbf{e}_K =
r_{00} $, for $ i=1,\cdots,N $. \item $ \mathbf{Y}_{0j} =
\mathbf{e}_K^T \mathbf{Y}_{ij} $ for $ i,j=1,\cdots,N $.
\end{enumerate}
\end{lem}
\begin{proof}
Noting $ \mathbf{TY}=\mathbf{0} $ (see Theorem \ref{thm:bary}.4),
the proof follows.
\end{proof}
If the Gangster operator is applied to (\ref{eq:SemiR}), it results
in the following redundant constraint $$ \textmd{diag}
(\mathbf{\hat{V}R \hat{V}}^T) = (1, (\mathbf{ \hat{V}R
\hat{V}}^T)_{0,1:n})^T. $$ Note that using Lemma \ref{lem:cons}, $
\mathbf{Y}_{0j} = \mathbf{e}_K^T \mathbf{Y}_{jj} $ for $
j=1,\cdots,N $ and the off-diagonal entries of each $
\mathbf{Y}_{jj} $ are zero. Therefore, by defining a new set $
\bar{J} = J \cup \left\lbrace 0 , 0 \right\rbrace $ and eliminating
the redundant constraints, we obtain a new SDP relaxation model
({\bf Model III}):
\begin{align}\label{eq:SDPF}
\nonumber
\min \;\; &\textmd{trace} (\mathbf{\hat{V}}^T\mathcal{L}_{\mathbf{Q}} \mathbf {\hat{V}}) \mathbf{R}\\
\nonumber
\textmd{s.t. }\; & \mathcal{G}_{\bar{J}}(\mathbf{\hat{V}R\hat{V}^T}) = {\bf E}_{00}\\
& \mathbf{R} \succeq 0,
\end{align}
where $ \mathbf{R} $ is an $ (N(K-1)+1) \times (N(K-1)+1) $ matrix
and $ {\bf E}_{00} $ is an $ (NK+1) \times (NK+1) $ all zero matrix
except for a single element equal to $ 1 $ in its $ (0,0) $th entry.
With this new index set $\bar{J}$, we are able to remove all the
redundant constraints while maintaining the SDP relaxation. The
relaxation model in (\ref{eq:SDPF}) corresponds to a tighter lower
bound and has an interior point in its feasible set as shown in the
following theorem.

\begin{thm}\label{thm:feasiblep}
The $ (N(K-1)+1) \times (N(K-1)+1) $ matrix
\begin{equation}
\hat {\bf R} = \left [
\begin{array}{c|c}
1 & \frac{1}{K} {\bf e}^T_{N(K-1)} \\[1.5ex]
\hline \frac{1}{K} {\bf e}_{N(K-1)} & \frac{1}{K^2}{\bf E}_{N(K-1)}
+ \frac{1}{K^2} {\bf I}_{N}\otimes (K{\bf I}_{K-1} - {\bf E}_{K-1} )
\end{array}
\right ]
\end{equation}
is a strictly interior point of the feasible set for the relaxation
problem (\ref{eq:SDPF}).
\end{thm}
\begin{proof}
The matrix $ \mathbf{\hat{R}} $ is positive definite. The rest of
the proof follows by showing $ \mathbf{ \hat{V} \hat{R} \hat{V}^T} =
\mathbf{\hat{Y}} $.
\end{proof}

%
The relaxation in (\ref{eq:SDPF}) is further tightened by
considering the \emph{non-negativity constraints} \cite{SotRen03}.
All the elements of the matrix $ \mathbf{Y} $ which are not covered
by the Gangster operator are greater than or equal to zero. These
inequalities can be added to the set of constraints in
(\ref{eq:SDPF}), resulting in a stronger relaxation model ({\bf
Model IV}):
\begin{align}\label{eq:SDPN}
\nonumber
\min \;\; &\textmd{trace} (\mathbf{\hat{V}}^T\mathcal{L}_{\mathbf{Q}} \mathbf {\hat{V}}) \mathbf{R}\\
\nonumber
\textmd{s.t. }\; & \mathcal{G}_{\bar{J}}(\mathbf{\hat{V}R\hat{V}^T}) = E_{00}\\
\nonumber & \mathcal{G}_{\hat{J}}(\mathbf{\hat{V}R\hat{V}^T}) \geq 0  \\
& \mathbf{R} \succeq 0,
\end{align}
where the set $ \hat{J} $ indicates those indices which are not
covered by $ \bar{J} $.

Note that this model is considerably stronger than model
(\ref{eq:SDPF}) because non-negativity constraints are also imposed
in the model. The advantage of this formulation is that the number
of inequalities can be adjusted to provide a trade-off between the
strength of the bounds and the complexity of the problem. The larger
number of the constraints in the model is, the better it
approximates the optimization problem (\ref{eq:semisum}) (with an
increase in the complexity).

The most common methods for solving SDP problems of moderate sizes
(with dimensions on the order of hundreds) are IPMs, whose
computational complexities are polynomial, see e.g. \cite{AlHaOv98}.
There are a large number of IPM-based solvers to handle SDP
problems, e.g., DSDP \cite{BenYe04}, SeDuMi \cite{Stu01}, SDPA
\cite{KoFuNaYa02}, etc. In our numerical experiments, we use DSDP
and SDPA for solving (\ref{eq:SDPF}), and SeDuMi is implemented for
solving (\ref{eq:SDPN}). Note that adding the non-negativity
constraints increases the computational complexity of the model.
Since the problem sizes of our interest are moderate, the complexity
of solving (\ref{eq:SDPN}) with IPM solvers is tractable.

\section{Randomization Method}\label{sec:rand}

Solving the SDP relaxation models (\ref{eq:SDPF}) and
(\ref{eq:SDPN}) results in a matrix ${\bf R}$. This matrix is
transformed to ${\bf Y}$ using ${\bf Y}={\hat{\bf V}}{\bf
R}{\hat{\bf V}}^T$, whose elements are between 0 and 1. This matrix
has to be converted to a binary rank-one solution of
(\ref{eq:semisum}), i.e. ${\bf Y}_{\bf u}$, or equivalently, a
binary vector ${\bf u}$ as a solution for (\ref{eq:semiprob}).

For any feasible point of (\ref{eq:semisum}), i.e. ${\bf Y}_{\bf
u}$, the first row, the first column, and the vector of the diagonal
elements of this symmetric matrix are equal to a binary solution for
(\ref{eq:semiprob}). For any matrix ${\bf Y}$ resulting from the
relaxation problems (\ref{eq:SDPF}) or (\ref{eq:SDPN}), its first
row, its first column, and the vector of its diagonal elements are
equal. Therefore, the vector ${\bf u}$ is approximated by rounding
off the elements of the first column of the matrix ${\bf Y}$.
However, this transformation results in a loose upper bound on the
performance.
In order to improve the performance, ${\bf Y}$ is transformed to a
binary rank-one matrix through a randomization procedure. An
intuitive explanation of the randomization procedure is presented in
\cite{StLuWo03}. We present two randomization algorithms in the
Appendix \ref{app:rand}.

\section{Complexity Reduction Using Lattice Basis Reduction}

Lattice structures have been used frequently in different
communication applications such as quantization or MIMO decoding. A
real lattice $ \Lambda $ is a discrete set of $M$-dimensional
vectors in the real Euclidean $ M $-space, $ \mathbb{R}^{M} $, that
forms a group under the ordinary vector addition. Every lattice $
\Lambda $ is generated by the integer linear combinations of a set
of linearly independent vectors $ \{ \mathbf{b}_{1},\cdots
,\mathbf{b}_{N}\} $, where $ {\bf b}_i \in \Lambda $, and the
integer $ N, N \leq M $, is called the dimension of the
lattice\footnote{Without loss of generality, we assume that $ N=M
$.}. The set of vectors $ \{ \mathbf{b}_{1},\cdots ,\mathbf{b}_{N}
\} $ is called a basis of $ \Lambda $, and the $ N \times M $ matrix
$ B = [\mathbf{b}_{1}, \cdots , \mathbf{b}_{N}]^{T} $ which has the
basis vectors as its rows is called the basis matrix (or generator
matrix) of $ \Lambda $.

The basis for representing a lattice is not unique. Usually a basis
consisting of relatively short and nearly orthogonal vectors is
desirable. The procedure of finding such a basis for a lattice is
called \textit{Lattice Basis Reduction}. Several distinct notions of
reduction have been studied, including Lenstra-Lenstra and
Lovasz (LLL) reduced basis \cite{LLL82}, which can be computed in 
polynomial time.

An initial solution for the lattice decoding problem can be computed
using one of the simple sub-optimal algorithms such as ZFD or
channel inversion, e.g. $ {\bf s}^\prime=\left[{\bf H}^{-1}{\bf
y}\right]$. If the channel is not ill-conditioned, i.e. the columns
of the channel matrix are nearly orthogonal and short, it is most
likely that the ML solution of the lattice decoding problem is
around ${\bf s}^\prime$. Therefore, using a reduced basis for the
lattice, each $ x_i $ in (\ref{eq:idear}) can be expressed by a few
points in $\mathcal{S}$ around $s^\prime_i$, not all the points in
$\mathcal{S}$. In general, this results in a sub-optimal algorithm.
However, for the special case of a MIMO system with two antennas
(with real coefficients), it has been shown that by using the LLL
approximation and considering two points per dimension we achieve
the ML decoding performance \cite{YaoWor02}.
%

Let $ \mathbf{L} = \mathbf{H}\mathbf{Q} $ be the LLL reduced basis
for the channel matrix $ \mathbf{H} $, where $ \mathbf{Q} $ is a
unimodular matrix. The MIMO system model in (\ref{eq:realchan}) can
be written as
\begin{equation}\label{eq:lllchan}
\mathbf{y} = \mathbf{L}\mathbf{Q}^{-1}\mathbf{x} + \mathbf{n}.
\end{equation}
Consider the QAM signaling. Without loss of generality, we can
assume coordinates of ${\bf x}$ are in the integer grid. Since ${\bf
Q}$ is a unimodular matrix, the coordinates of a new variable
defined as $ \mathbf{x}^{\prime} = \mathbf{Q}^{-1} \mathbf{x} $ are
also in the integer grid. Therefore, the system in
(\ref{eq:lllchan}) is modelled by $ \mathbf{y} = \mathbf{L}
\mathbf{x}^{\prime} + \mathbf{n} $. Note that by multiplying ${\bf
x}$ by ${\bf Q}^{-1}$ the constellation boundary will change.
However, it is shown that in the lattice decoding problem with
finite constellations the best approach is to ignore the boundary
and compute the solution \cite{DaElCa04}. If the solution is outside
the region, it is considered as an error. This change of boundary
will result in some performance degradation. The performance
degradation for some scenarios are depicted in Fig. \ref{fig:ber4}
and Fig. \ref{fig:ber6}.

In order to implement the proposed method using LLL basis reduction,
each component of ${\bf x}^\prime$ is expressed by a linear
combination (with zero-one coefficients) of $L$ (usually much
smaller than $K$) integers around $s^\prime_i$, where ${\bf
s}^\prime = \left[{\bf L}^{-1}{\bf y}\right]$. Then, the proposed
algorithm can be applied to this new model. Due to the change of
constellation boundary, there is a degradation in the performance.
However, the complexity reduction is large. The trade-off between
performance degradation and complexity reduction can be controlled
by the choice of $L$ (see simulation results). The reduction in the
complexity is more pronounced for larger constellations. Note that
the dimension of the semi-definite matrix $ \mathbf{Y} $ is $
N*(K-1) + 1 $. Therefore, the LLL reduction decreases the dimension
of the matrix $ \mathbf{Y} $ to $ N*(L-1) + 1 $ (where usually $L
\ll K$), and consequently, decreases the computational complexity of
the proposed algorithm. The performance of this method is shown in
the simulation results.

\section{Extension for Soft Decoding}

%
In this section, we extend our proposed SDP relaxation decoding
method for soft decoding in MIMO systems. The SDP soft decoder is
derived as an efficient solution of the max-approximated soft ML
decoder. The complexity of this method is much less than that in the
soft ML decoder. Moreover, the performance of the proposed method is
comparable with that in the ML one. Also, the proposed method can be
applied to any arbitrary constellation and labelling method, say
Grey labeling.

In the MIMO system defined in (\ref{eq:realchan}), any transmit data
$ {\bf x} $ is represented by $ N_b = \log_2 K $ bits ($ {\bf x} =
{\rm map}({\bf b}) $, where ${\bf b}$ is the corresponding binary
input). Given a received vector $ \mathbf{y} $, the soft decoder
returns the soft information about the likelihood of $ b_j = 0 ~{\rm
or}~ 1, ~j=1,\cdots,NN_b $. The likelihoods are calculated by
Log-Likelihood Ratios (LLR) in a Maximum A Posteriori (MAP) decoder
by
\begin{equation}\label{eq:llr}
\mathcal{L}(b_j|{\bf y}) = \log \left( \dfrac{P(b_j=1|{\bf
y})}{P(b_j=0|{\bf y})} \right).
\end{equation}
Define
\begin{equation}\label{eq:llrA}
\mathcal{L}_{A}(b_j|{\bf y}) = \log \dfrac{P(b_j=1)}{P(b_j=0)}.
\end{equation}
It is shown that the LLR values are formulated by \cite{HocBri03}
\begin{eqnarray}\label{eq:llrc}
\nonumber\mathcal{L}(b_j|{\bf y}) &=& \underbrace{ \log
\dfrac{\sum_{{\bf b} \in \mathbb{B}_{k,1}}p({\bf y}|{\bf b}).
\exp\left(\frac{1}{2}{\bf b}_{[k]}^T.{\bf L}_{A,[k]} \right)}
{\sum_{{\bf b} \in \mathbb{B}_{k,0}}p({\bf y}|{\bf b}). \exp \left(
\frac{1}{2}{\bf b}_{[k]}^T.{\bf
L}_{A,[k]}\right)}}_{\mathcal{L}_E(b_k|{\bf y})}\\
&+& \mathcal{L}_{A}(b_j|{\bf y}),
\end{eqnarray}
where $ {\bf b}_{[k]} $ denotes the sub-vector of $ {\bf b} $
obtained by omitting its $k$th element $ b_k $, $ {\bf L}_{A,[k]} $
denotes the vector of all $ \mathcal{L}_{A} $ values, also omitting
$ b_k $, and $ \mathbb{B}_{k,1} $ (resp. $ \mathbb{B}_{k,0} $)
denotes the set of all input vectors, $ {\bf b} $, such that $ b_k =
1 $ (resp. $ b_k = 0 $). Note that there is an isomorphism between $
\mathbb{B}_{k,1} $ (resp. $ \mathbb{B}_{k,0} $) and $
\mathbb{X}_{k,1} $ (resp. $ \mathbb{X}_{k,0} $), where $
\mathbb{X}_{k,1} $ (resp. $ \mathbb{X}_{k,0} $) denotes the set of
all corresponding constellation symbols, $ \mathbb{X}_{k,1}= \left
\lbrace {\bf x} : {\bf x} = {\rm map} ({\bf b}) , {\bf b} \in
\mathbb{B}_{k,1} \right \rbrace $ (resp. $ \mathbb{X}_{k,0}= \left
\lbrace {\bf x} : {\bf x} = {\rm map} ({\bf b}) , {\bf b} \in
\mathbb{B}_{k,0} \right \rbrace $).

As shown in \cite{HocBri03}, the computation of the LLR values in
(\ref{eq:llrc}) requires computing the likelihood function $ p({\bf
y}|{\bf b})$, i.e.
\begin{equation}\label{eq:llrmap}
p({\bf y}|{\bf x}={\rm map}({\bf b}))=\dfrac{\exp\left[
-\frac{1}{2\sigma^2}.\parallel{\bf y} - {\bf Hx} \parallel^2
\right]}{\left(2\pi\sigma^2\right)^N},
\end{equation}
where $ \sigma^2 = \frac{1}{SNR} $.

By having the likelihood functions, these LLR values are
approximated efficiently using the Max-log approximation
\cite{HocBri03}
\begin{eqnarray}\label{eq:llrmax}
\nonumber \mathcal{L}_E(b_k|{\bf y})\approx
\hspace{-9pt}&+&\hspace{-9pt}\frac{1}{2} \max_{{\bf b} \in
\mathbb{B}_{k,1}} \hspace{-3pt}\left\lbrace \hspace{-3pt}
-\frac{1}{\sigma^2} \parallel {\bf y} \hspace{-2pt}-\hspace{-2pt}
{\bf Hx} \parallel^2 +
{\bf b}_{[k]}^T.{\bf L}_{A,[k]} \hspace{-3pt}\right\rbrace\\
\hspace{-9pt}&-&\hspace{-9pt} \frac{1}{2} \max_{{\bf b} \in
\mathbb{B}_{k,0}} \hspace{-3pt}\left\lbrace \hspace{-3pt}
+\frac{1}{\sigma^2} \parallel {\bf y} \hspace{-2pt}-\hspace{-2pt}
{\bf Hx} \parallel^2 + {\bf b}_{[k]}^T.{\bf L}_{A,[k]}
\hspace{-3pt}\right\rbrace.
\end{eqnarray}

Without loss of generality, we assume that all components, $ x_i $,
of an input vector are equiprobable\footnote{In order to consider
the effects of non-equiprobable symbols, both approaches presented
in \cite{StLuWo031} and \cite{WanGia04} can be applied.}; therefore,
the second term in each maximization in (\ref{eq:llrmax}) will be
removed. Hence, computing the LLR values requires to solve problems
of the form
\begin{equation}\label{eq:softdec}
\min_{{\bf x} \in \mathbb{X}_{k,\zeta}} \parallel {\bf y} - {\bf Hx}
\parallel^2,
\end{equation}
where $ k=1,\cdots,NN_b $ and $ \zeta =0 \; {\rm or} \; 1 $. Note
that, as mentioned in \cite{WanGia04}, only $ NN_b+1 $ problems
among $ 2NN_b $ problems of the form (\ref{eq:softdec}) are
considered.

The Quasi-Maximum likelihood decoding method proposed in this paper
can be applied to the problem (\ref{eq:softdec}). However, $
\mathbb{X}_{k,\zeta} $ must be defined in implementing the
algorithm. This set includes all the input vectors, $ {\bf x} \in
\mathcal{S}^N $, such that $ b_k=\zeta $. Assigning 0 or 1 to one of
the bits in $ {\bf b} $ removes half of the points in $
\mathcal{S}^N $. In other words, when $ b_k = \zeta $, one of the
components of the input vector $ {\bf x} $, say $ x_p $, can only
select half of the points in the set $ \mathcal{S} $, say $ \left
\lbrace s_{p_1},\cdots, s_{p_{\frac{K}{2}}}\right \rbrace $.
Therefore, the $p$th component of ${\bf x}$ is represented by
\begin{equation}\label{eq:ideasoft}
x_p=u_p(1)s_{p_1}+\cdots+u_p(\frac{K}{2})s_{p_{\frac{K}{2}}}.
\end{equation}

As a result, we have the same matrix expression $ {\bf x = S u} $ as
(\ref{eq:idear}), except that the length of the vector $ {\bf u} $
is $ (N-1)*K+\frac{K}{2} $ and in the $p$th row of the matrix $ {\bf
S} $, we have $ \frac{K}{2} $ elements $ \left \lbrace
s_{p_1},\cdots, s_{p_{\frac{K}{2}}} \right \rbrace $, instead of $K$
elements $ \left\lbrace s_1,\cdots,s_K \right \rbrace $. Now, the
proposed method can be applied to the new equation based on the new
matrix $ {\bf S} $ and $ {\bf u} $.

\section{Simulation Results}

\subsection{Performance Analysis}

\begin{figure}[htbp]
\centering
\includegraphics[scale=0.6]{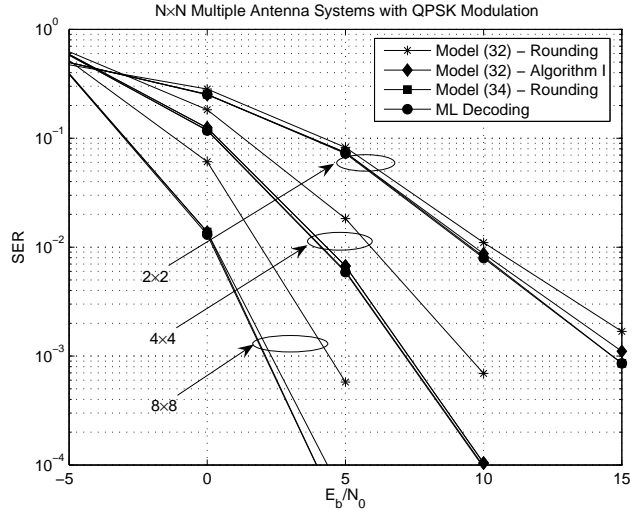}\\
\caption{Performance of the proposed model (\ref{eq:SDPF}) and
(\ref{eq:SDPN}) in a MIMO system with $N$ transmit and $N$ receive
antennas employing QPSK} \label{fig:ber1}
\end{figure}

\begin{figure}[htbp]
\centering
\includegraphics[scale=0.6]{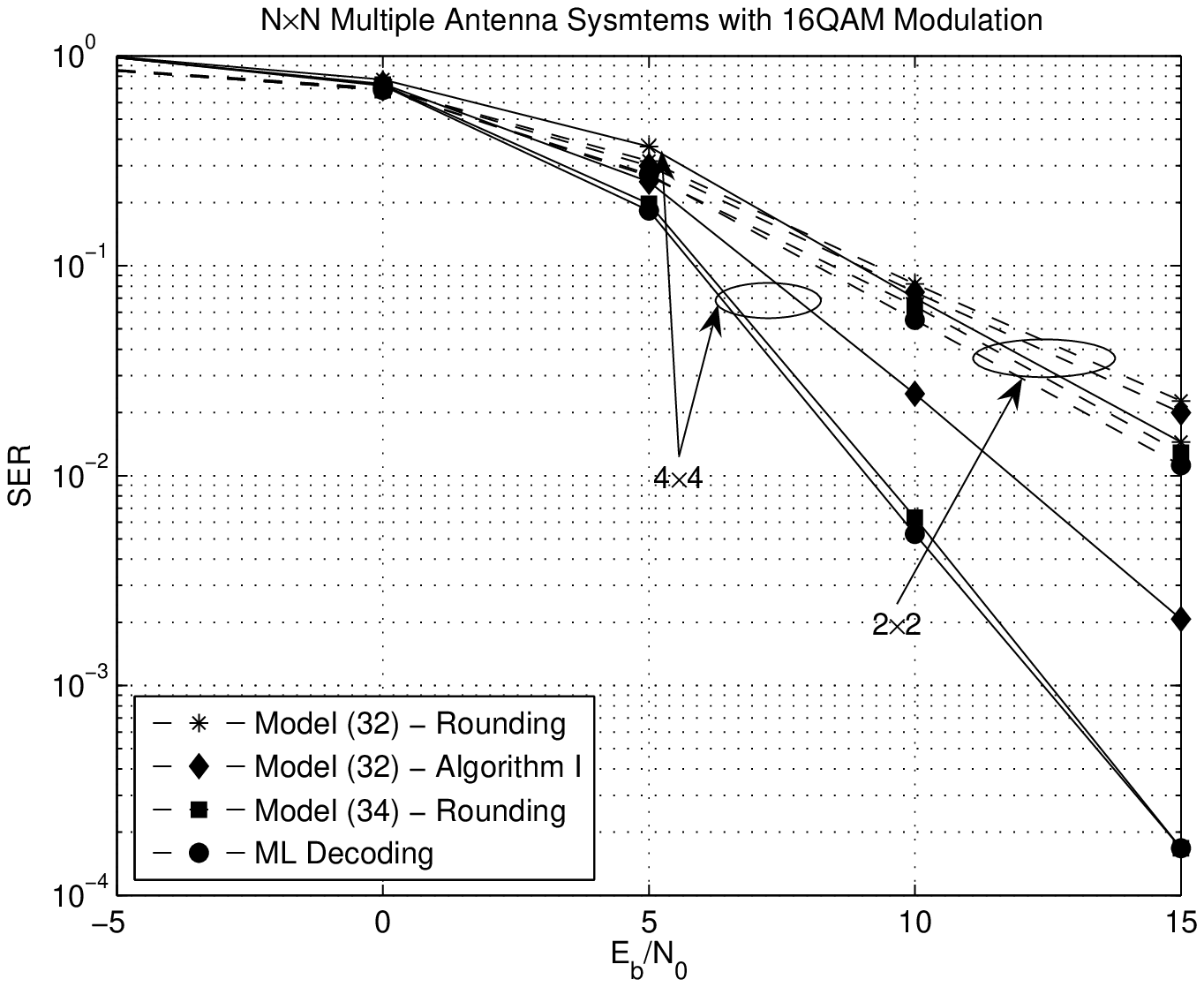}\\
\caption{Performance of the proposed model (\ref{eq:SDPF}) and
(\ref{eq:SDPN}) in a MIMO system with $N$ transmit and $N$ receive
antennas employing 16-QAM} \label{fig:ber2}
\end{figure}

\begin{figure}[htbp]
\centering
\includegraphics[scale=0.6]{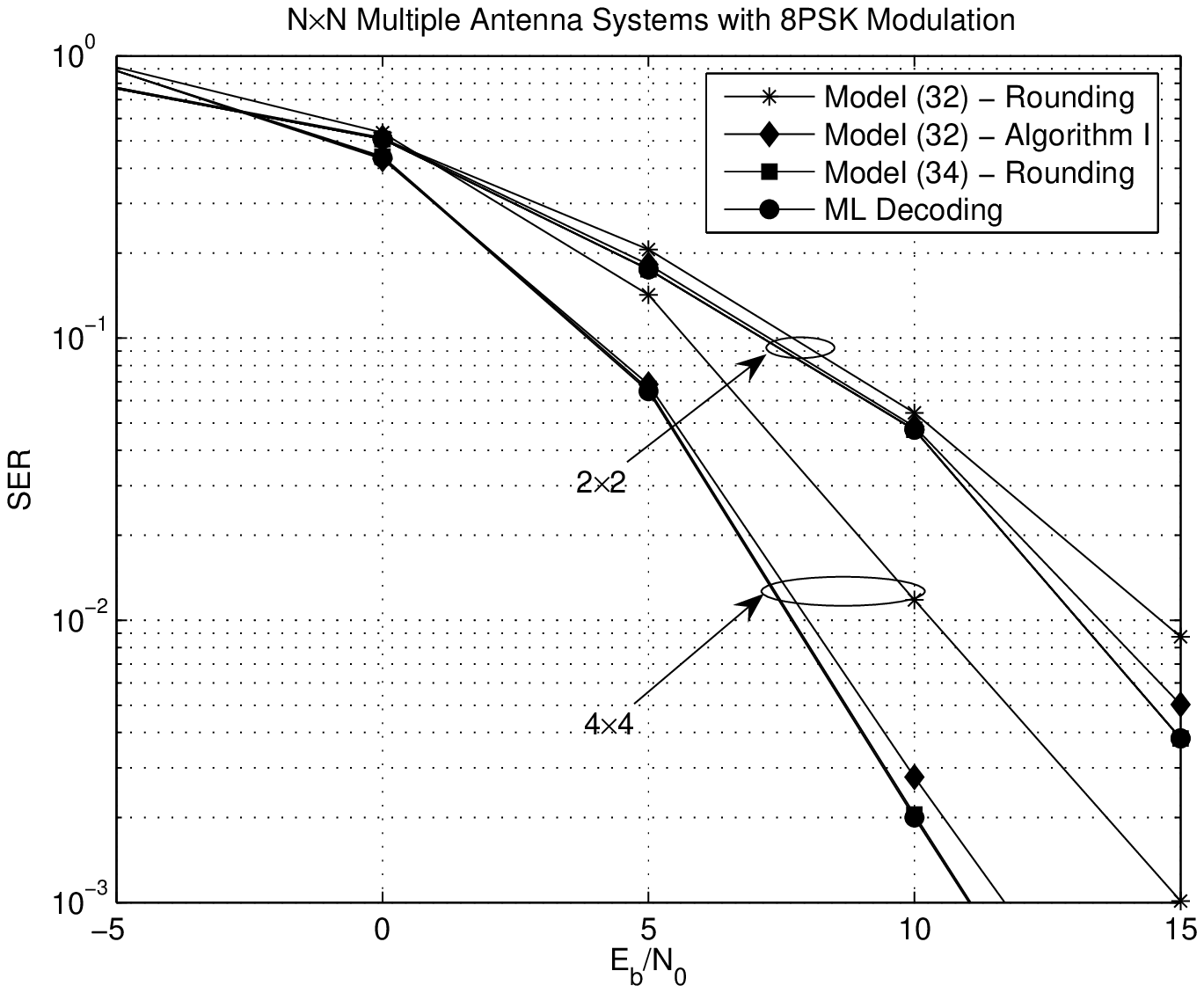}\\
\caption{Performance of the proposed model (\ref{eq:SDPF}) and
(\ref{eq:SDPN}) in a MIMO system with $N$ transmit and $N$ receive
antennas employing 8-PSK} \label{fig:ber5}
\end{figure}

We simulate the two proposed models (\ref{eq:SDPF}) and
(\ref{eq:SDPN}) for decoding in MIMO systems with QAM and PSK
constellations. Fig. \ref{fig:ber1} demonstrates that the proposed
quasi-ML method using model (\ref{eq:SDPF}) and the randomization
procedure achieves near ML performance in an un-coded $2\times 2$
MIMO system with QPSK constellation. Fig. \ref{fig:ber2} shows the
performance in a $4\times 4$ MIMO system with 16-QAM. The
performance analysis of a MIMO system with different number of
antennas employing $8$-PSK is shown in Fig. \ref{fig:ber5}. In
figures \ref{fig:ber1}, \ref{fig:ber2}, and \ref{fig:ber5}, the
curved lines with the stars represent the performance of the system
using relaxation model (\ref{eq:SDPF}), while a simple rounding
algorithm, as described in Section \ref{sec:rand}, transforms matrix
${\bf Y}$ to the binary vector ${\bf u}$. The ML decoding
performance is also denoted by a curved line with circles. By
increasing the dimension, the resulting gap between the relaxation
model (\ref{eq:SDPF}) and the ML decoding increases. However, using
the randomization Algorithm I with $M_{rand}=30$ to $50$
significantly decreases this gap (curved line with diamonds). The
curved lines with squares show the performance of the relaxation
model (\ref{eq:SDPN}) with a simple rounding, in which all the
non-negative constraints are included. This curve is close to ML
performance. It is clear that the relaxation model (\ref{eq:SDPN})
is much stronger than the relaxation model (\ref{eq:SDPF}). Note
that adopting different number of non-negative constraints will
change the performance of the system between the two curves with
diamonds and squares. In other words, the trade-off between
complexity and performance relies on the number of extra
non-negative constraints.

\begin{figure}[htbp]
\centering
\includegraphics[scale=0.6]{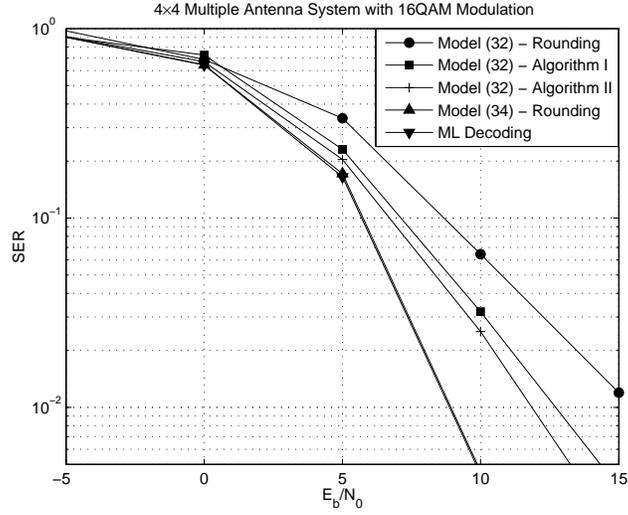}\\
\caption{Different randomization algorithms in a MIMO system with
$4$ transmit and $4$ receive antennas employing 16-QAM}
\label{fig:ber3}
\end{figure}

Fig. \ref{fig:ber3} compares the two proposed Randomization
procedure for the relaxation model (\ref{eq:SDPF}) and
(\ref{eq:SDPN}). The effect of the randomization methods, Algorithm
I and II, for the relaxation model (\ref{eq:SDPF}) is shown. As
expected, Algorithm II performs slightly better, while its
computational complexity is lower. The solution of the relaxation
model in (\ref{eq:SDPN}), in most cases, corresponds to the optimal
solution of the original problem (\ref{eq:semiprob}). In the other
words, because the model in (\ref{eq:SDPN}) is strong enough, there
is no need for the randomization algorithm. Several compromises for
improving the performance can be done, e.g. including only some of
the non-negative constraints in (\ref{eq:SDPN}) and/or using a
randomization procedure with a fewer number of iterations.

\begin{figure}[htbp]
\centering
\includegraphics[scale=0.6]{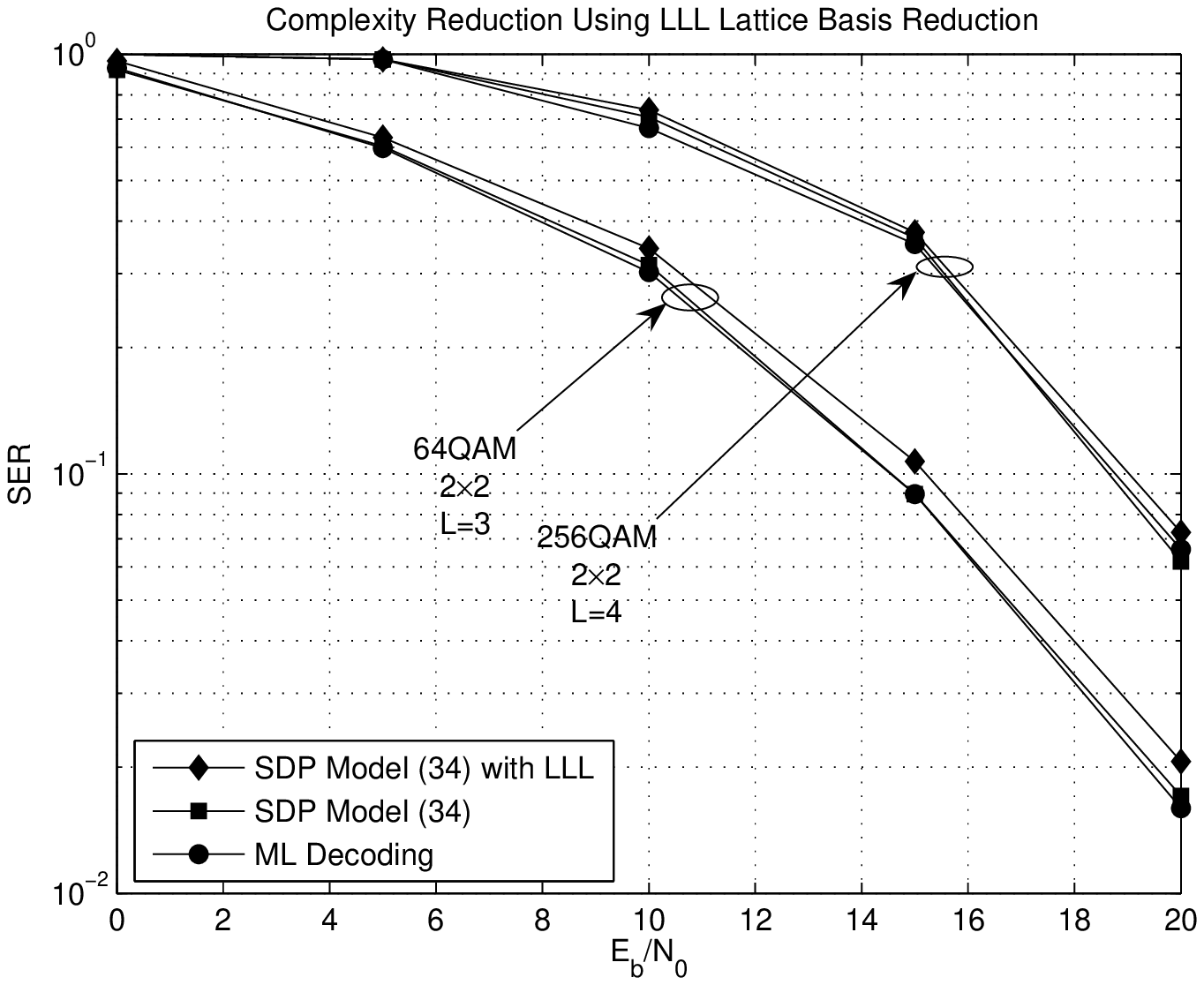}\\
\caption{Performance of using LLL lattice basis reduction for
relaxation model (\ref{eq:SDPN}) in a $2\times 2$ MIMO system with
$L=\log2(K)$} \label{fig:ber4}
\end{figure}

\begin{figure}[htbp]
\centering
\includegraphics[scale=0.6]{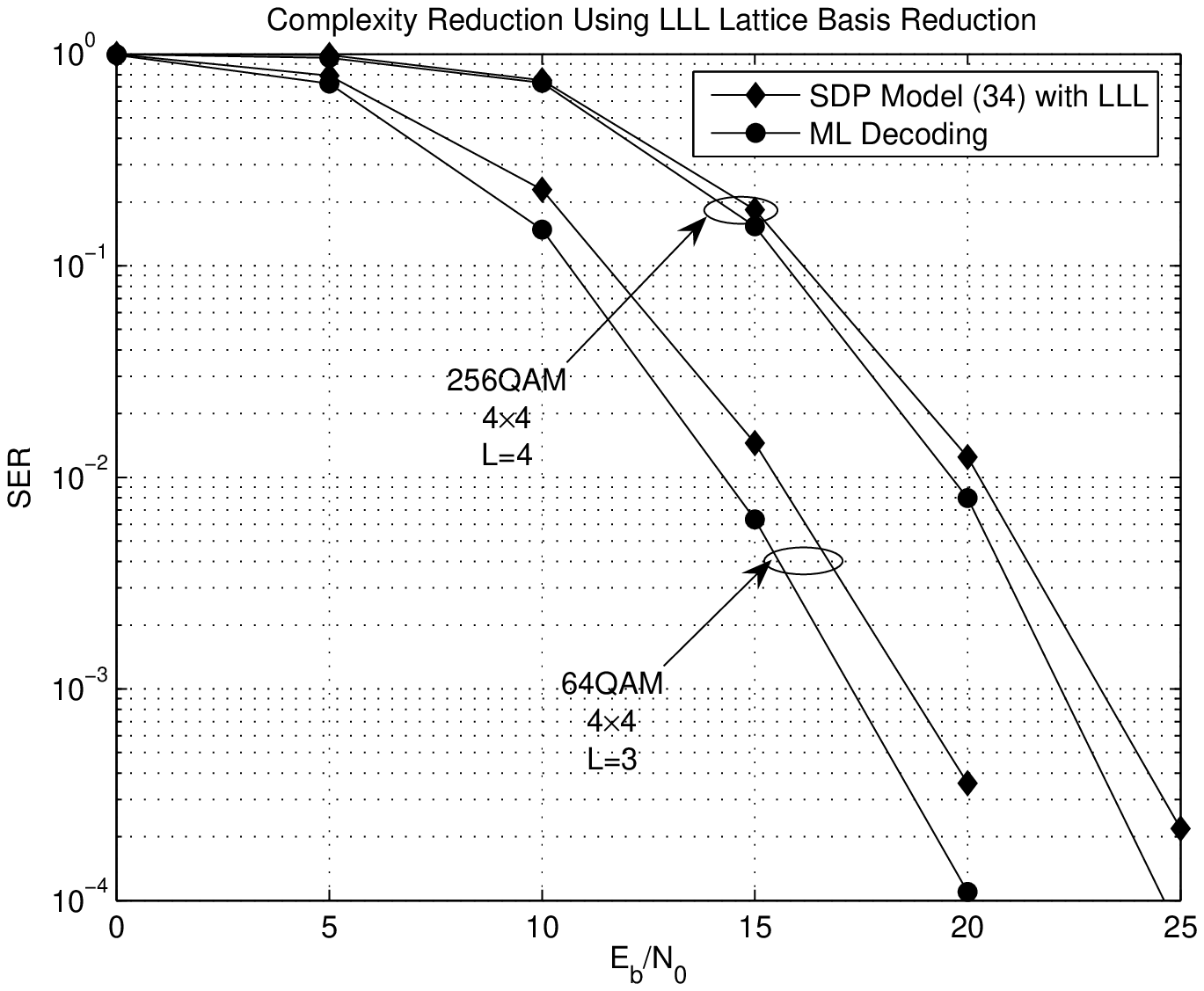}\\
\caption{Performance of using LLL lattice basis reduction for
relaxation model (\ref{eq:SDPN}) in a $4\times 4$ MIMO system with
$L=\log2(K)$} \label{fig:ber6}
\end{figure}

In order to reduce the computational complexity of the proposed
method, the LLL lattice basis reduction is implemented as a
pre-processing step for the relaxation model (\ref{eq:SDPN}). Fig.
\ref{fig:ber4} and Fig. \ref{fig:ber6} show the effect of using the
LLL lattice basis reduction in $2\times 2$ and $4\times 4$ multiple
antenna systems with $64$-QAM and $256$-QAM. In a system with
$64$-QAM and $256$-QAM, the performance of the relaxation model
(\ref{eq:SDPN}) is close to the ML performance with $K=8$ and
$K=16$, respectively. By using LLL reduction and considering
$L=\log_2(K)$ symbols around the initial point, the performance
degradation is acceptable, see Fig. \ref{fig:ber4} and Fig.
\ref{fig:ber6}. Note that the resulting gap in the performance is
small, while the reduction in computational complexity is
substantial.


\subsection{Complexity Analysis}

Semi-definite programs of reasonable size can be solved in
polynomial time within any specified accuracy by IPMs. IPMs are
iterative algorithms which use a Newton-like method to generate
search directions to find an approximate solution to the nonlinear
system. The IPMs converge vary fast and an approximately optimal
solution is obtained within a polynomial number of iterations. For a
survey on IPMs see \cite{Kle02,Ye97}. In the sequel, we provide an
analysis for the worst case complexity of solving models
(\ref{eq:SDPF}) and (\ref{eq:SDPN}) by IPMs.

It is known (see e.g.~\cite{Tod01}) that a SDP with rational data
can be solved, within a tolerance $\epsilon$, in $O(\sqrt{m}
\log(1/\epsilon))$ iterations, where $m$ is the dimension of the
matrix variable. Note that for the SDP problems (\ref{eq:SDPF}) and
(\ref{eq:SDPN}), $m=N(K-1)+1$.

The computational complexity for one interior-point iteration
depends on several factors. The main computational task in each
iteration is solving a linear system an order determined by the
number of constraints, $p$. This task requires $O(p^3)$ operations.
The remaining computational tasks involved in one interior-point
iteration include forming system matrix whose total construction
requires $O(pm^3+p^2m^2)$ arithmetic operations. Thus, the
complexity per iteration of the IPM for solving SDP problem whose
matrix variable is of dimension $m$ and number of equality
constraints $p$, is $O (pm^3+p^2m^2+p^3)$. This means for a given
accuracy $\epsilon$, an interior-point method in total requires at
most $O (p(m^3+pm^2+p^2 )\sqrt{m}$ $\log(1/\epsilon) )$ arithmetic
operations.

Since the SDP relaxation (\ref{eq:SDPF}) contains $O(K^2N)$ equality
constraints, it follows that a solution to (\ref{eq:SDPF}) can be
found in at most $O(N^{4.5}K^{6.5}\log(1/\epsilon))$ arithmetic
operations. SDP relaxation (\ref{eq:SDPN}) contains $O(K^2N)$
equations and $O(K^2N^2)$ sign constraints. In order to solve
relaxation (\ref{eq:SDPN}), we formulate the SDP model as a standard
linear cone program (see e.g.~\cite{Stu02}) by adding some slack
variables.
The additional inequality constraints make the model in
(\ref{eq:SDPN}) considerably stronger than the model in
(\ref{eq:SDPF}) (see numerical results),  but also  more difficult
to solve. An interior-point method for solving SDP model
(\ref{eq:SDPN}) within a tolerance $\epsilon$ requires at most
$O(N^{6.5}K^{6.5}\log(1/\epsilon))$ arithmetic operations. Since
the problem sizes of interest are moderate
, the problem in (\ref{eq:SDPN}) is tractable. However, there exist
a trade-off between the strength of the bounds and the computational
complexity for solving these two models (see Section
\ref{sec:semi}).

The complexity of the randomization procedure applied to the model
(\ref{eq:SDPF}) is negligible compared to that of solving the
problem itself. Namely, if we denote the number of randomization
iterations by $N_{rand}$, then the worst case complexity of the
randomization procedure is $O(NKN_{rand})$.

The problems (\ref{eq:SDPF}) and (\ref{eq:SDPN}) are polynomially
solvable. These problems have many variables; however, they contain
sparse low-rank (rank-one) constraint matrices. Exploiting the
structure and sparsity characteristic of semi-definite programs is
crucial to reduce the complexity. In \cite{BeYeZh99}, it is shown
that rank-one constraint matrices (similar to our problems) reduce
the complexity of the interior-point algorithm for positive
semi-definite programming by a factor of $NK$. In other words, the
complexities of the SDP relaxation problems (\ref{eq:SDPF}) and
(\ref{eq:SDPN}) are decreased to $O(N^{3.5}K^{5.5}\log(1/\epsilon))$
and $O(N^{5.5}K^{5.5}\log(1/\epsilon))$, respectively. Also,
implementing the rank-one constraint matrices results in a faster
convergence and a saving in the computation time and memory
requirements. It is worth mentioning that when we use the LLL
lattice basis reduction, the value of $K$ is replaced with $L$ in
the aforementioned analysis. As mentioned before, this value is much
smaller than $K$, e.g. in our simulation results $L=\log_2(K)$,
which results in reducing the computational complexity.

\subsection{Comparison}

The \textit{worst-case complexity} of well-known SD method
\cite{HasVik031, AgVaZe02} is known to be an exponential function of
dimension $M$ over all ranges of rate and $S\!N\!R$
\cite{HasVik031}. The complexity analysis shows that our proposed
SDP algorithms possess a polynomial-time worst case complexity. It
should be emphasized that in real time problems, the time spent for
decoding the received vector is important and it can be considered
as a measure of the complexity.

In the following, the worst case complexities of the algorithm based
on model \eqref{eq:SDPF}, the method proposed in \cite{WiElSh05L},
the method in \cite{StLuWo03}, and the SD algorithm \cite{AgVaZe02}
are compared with different random values of input vector, channel
matrix, and noise for $E_b/N_0=\{-5,0,5,10,15\}$. For each value of
$E_b/N_0$, the algorithms are performed for $10^5$ times and the
maximum time spent for the decoding procedure is saved in $MaxTime$.
The average time spent for decoding each case is stored in
$AveTime$.

It should be emphasized that the $MaxTime$ for each case depends on
how the algorithm is implemented. There are numerous variants for SD
algorithm. In the following, we have implemented the SD algorithm
based on the Schnorr-Euchner strategy proposed in \cite{AgVaZe02}.
Moreover, the simulations of the proposed algorithms are implemented
by one of the simplest available packages, the SDPA package
\cite{KoFuNaYa02}. However, by utilizing the sparsity of the
constraint matrices as suggested in \cite{BeYeZh99} and using the
DSDP package, the computed $AveTime$ and $MaxTime$ can be reduced
dramatically (a factor of $NK$ in the analysis), without any
performance degradation.

Table \ref{Tab:Max4_4_4} shows the simulation results for a MIMO
system with $\tilde{M}=\tilde{N}=4$ employing 16-QAM. The maximum
time for decoding a symbol using SD algorithm is much longer than
the corresponding time in the proposed SDP relaxation method. The
other three methods have comparable $MaxTime$. As it is also shown
in the analysis, the proposed Model III is more complex compared to
the two other SDP methods. However, this method outperforms the
other SDP methods in \cite{StLuWo03} and \cite{WiElSh05L}.

\begin{table}
  \centering
  \caption{Comparison of $MaxTime$ for different methods in a MIMO system with 4 transmit and 4 receive antennas
  employing 16-QAM}\label{Tab:Max4_4_4}
  \begin{tabular}{|l|ccccc|}
  \hline
  $E_b/N_0$ & -5 & 0 & 5 & 10 & 15\\
  \hline
  Model III & 0.1037  &  0.1095  &  0.1108  &  0.1178  &  0.1196\\
  Method \cite{WiElSh05L} & 0.0685 &   0.0640  &  0.0697  &  0.0735  &  0.0624\\
  Method \cite{StLuWo03} & 0.0580  &  0.0633  &  0.0536  &  0.0646  &  0.0596\\
  SD Method & 61.8835  & 47.0480 &  28.0347  &  4.3848  &  2.2477\\
  \hline
  \end{tabular}
\end{table}

\begin{figure}[htbp]
\centering
\includegraphics[scale=0.6]{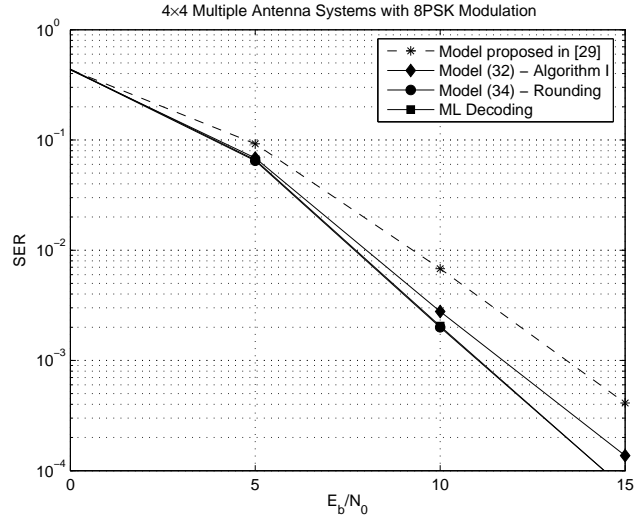}\\
\caption{Comparison of the relaxation model proposed in
\cite{LuLuKi03} and that in our method in a MIMO system with 4
transmit and receive antennas, employing 8-PSK modulation}
\label{fig:comp1}
\end{figure}

\begin{figure}[htbp]
\centering
\includegraphics[scale=0.6]{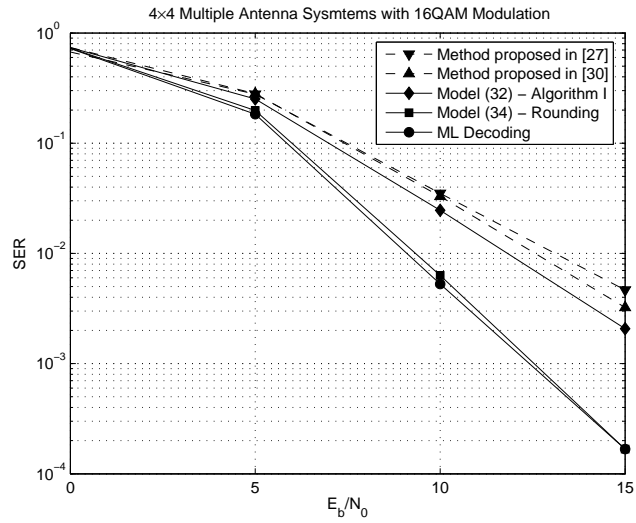}\\
\caption{Comparison of the relaxation model proposed in
\cite{WiElSh05L} and that in our method in a MIMO system with 4
transmit and receive antennas, employing 16-QAM modulation}
\label{fig:comp2}
\end{figure}

The relaxation model (\ref{eq:SDPF}) outperforms the SDP methods
proposed in \cite{StLuWo03} and \cite{LuLuKi03}. Fig.
\ref{fig:comp1} compares the performance of \cite{LuLuKi03} and the
relaxation model \eqref{eq:SDPF} and the performance of the method
proposed in \cite{StLuWo03} is shown in Fig. \ref{fig:comp2} in a
MIMO system with 4 transmit and receive antennas. The order of the
complexity of \cite{StLuWo03} is comparable to the proposed model
(\ref{eq:SDPF}) and the order of the complexity of \cite{LuLuKi03}
is less than that of the model (\ref{eq:SDPF}) ($O(N^2)$ vs.
$O(N^{3.5})$). The method in \cite{StLuWo03} can handle QAM
constellations; however, it achieves near ML performance only in the
case of BPSK and QPSK constellations. Also, the method in
\cite{LuLuKi03} is limited to PSK constellations. Note that our
proposed methods can be used for any arbitrary constellation and
labeling.

The comparison of the performance of the relaxation model proposed
in \cite{WiElSh05L} and that of our method is shown in Fig.
\ref{fig:comp2} ($4 \times 4$ antenna system employing $16$-QAM). It
is observed that the SDP relaxation models (\ref{eq:SDPF}) and
(\ref{eq:SDPN}) perform better than \cite{WiElSh05L}. The order of
the complexity of \cite{WiElSh05L} is the same as that of the model
(\ref{eq:SemiR}), while the model (\ref{eq:SDPN}) is more complex
($O(N^{5.5})$ vs. $O(N^{3.5})$).


Although the worst case complexities of the SD algorithm
\cite{AgVaZe02, HasVik031} and the other variants are exponential,
in several papers, the average complexity of these algorithms are
investigated. In \cite{JalOtt05}, it is shown that generally, there
is an exponential lower bound on the \textit{average complexity} of
the SD algorithm. However, it is shown that for large values of
$E_b/N_0$ and small values of dimension $M$, the average complexity
can be approximated by a polynomial function of dimension $M$.

In Table \ref{Tab:Ave4_4_4}, the average time $AveTime$ spent for
decoding the received vectors in the previous scenario is shown. As
it can be seen, the average complexity of all SDP methods is
gradually increasing with $E_b/N_0$ while the average complexity of
SD method is decreasing exponentially. This suggests that for
different dimensions $M$ and values of $E_b/N_0$, there is a
threshold that the proposed SDP methods perform better than SD
algorithm even in terms of the average complexity. However, Table
\ref{Tab:Max4_4_4} shows that how inefficient SD algorithm performs
in terms of the worst-case complexity.

\begin{table}
  \centering
  \caption{Comparison of $AveTime$ for different methods in a MIMO system with 4 transmit and 4 receive antennas
  employing 16QAM}\label{Tab:Ave4_4_4}
  \begin{tabular}{|l|ccccc|}
  \hline
  $E_b/N_0$ & -5 & 0 & 5 & 10 & 15\\
  \hline
  Model III & 0.0372  &  0.0377  &  0.0394  &  0.0428  &  0.0417\\
  Method \cite{WiElSh05L} &  0.0130  &  0.0134  &  0.0142  &  0.0156  &  0.0156\\
  Method \cite{StLuWo03} & 0.0116  &  0.0118  &  0.0126  &  0.0141  &  0.0141\\
  SD Method & 0.0449  &  0.0139  &  0.0060  &  0.0026  &  0.0016\\
  \hline
  \end{tabular}
\end{table}

In the following, in Tables \ref{Tab:4_4_2} and \ref{Tab:8_8}, the
performance of the proposed algorithm based on Model III and SD
algorithm are shown in terms of $AveTime$ and $MaxTime$, for
different number of antennas and constellations. It can be seen
that, in terms of the worst-case complexity the proposed algorithm
based on Model III always outperforms SD algorithm. Generally, we
can conclude that by increasing the dimension and rate, the range of
$E_b/N_0$ that the proposed model outperforms the SD algorithm
increases. In order to show that the $MaxTime$ values are not
sporadic, we also provide the values of $AveMaxTime$ in Table
\ref{Tab:8_8}. This number is the average of the largest $100$
decoding times in each case.

\begin{table}
  \centering
  \caption{Decoding Time in a MIMO system with 4 transmit and 4 receive antennas
  employing QPSK}\label{Tab:4_4_2}
  \begin{tabular}{|l|l|ccccc|}
  \cline{2-7}
  \multicolumn{1}{c|}{}& $E_b/N_0$ & -5 & 0 & 5 & 10 & 15\\
  \hline
  \multirow{2}{*}{$AveTime$} & Model III & 0.0154  &  0.0156  &  0.0238  &  0.0278  &  0.0236\\
  & SD Method & 0.0199  &  0.0074  &  0.0046  &  0.0028  &  0.0020\\
  \hline
  \multirow{2}{*}{$MaxTime$} & Model III & 0.4271  &  0.4251  &  0.4765  &  0.7572  &  0.8417\\
  & SD Method & 28.326  & 26.3109  & 25.4260   & 2.2232 &   0.9663\\
  \hline
  \end{tabular}
\end{table}

\begin{table}
  \centering
  \caption{Decoding Time in a MIMO system with 8 transmit and 8 receive antennas}
  \label{Tab:8_8}
  \begin{tabular}{|c|l|l|ccccc|}
  \cline{3-8}
  \multicolumn{2}{c|}{}& $E_b/N_0$ & -5 & 0 & 5 & 10 & 15\\
  \hline
  \multirow{6}{*}{QPSK} & \multirow{2}{*}{$AveTime$} & Model III &
  0.0152 & 0.0152 & 0.0174 & 0.0224 & 0.0306\\
  & & SD Method & 0.6005 &   0.1061  &  0.0319  &  0.0149  &  0.0052\\
  \cline{2-8}
  & \multirow{2}{*}{$MaxTime$} & Model III &  0.0965 & 0.0655 & 0.1666 &
  0.6586 & 0.6959\\
  & & SD Method & 433.3972 & 179.0310  & 19.7889  & 16.7787   & 7.7819\\
  \cline{2-8}
  & \multirow{2}{*}{$AveMaxTime$} & Model III & 0.0658 & 0.0587 & 0.0642 &
  0.1492 & 0.2109\\
  & & SD Method & 73.4830 & 16.3274  & 6.1652  & 5.9249   & 1.8074\\
  \hline
  \multirow{6}{*}{16-QAM} & \multirow{2}{*}{$AveTime$} & Model III &
  0.0936 & 0.0948 & 0.0984 & 0.1050 & 0.1059\\
  & & SD Method &  42.2894 &   1.6575   & 0.4762 &   0.2955  &  0.1080\\
  \cline{2-8}
  & \multirow{2}{*}{$MaxTime$} & Model III & 0.2867 & 0.2974 &
  0.2772 & 0.2916 & 0.3273\\
  & & SD Method &  8633.8 &  383.40  &  290.89  &  121.92  &  91.032\\
  \cline{2-8}
  & \multirow{2}{*}{$AveMaxTime$} & Model III & 0.1574 & 0.1580 & 0.1633 &
  0.1682 & 0.1712\\
  & & SD Method & 411.6743  & 15.3724  &  4.5987  &  2.9336  &  1.0162\\
  \hline
  \end{tabular}
\end{table}

The performance of the proposed SDP relaxation model
(\ref{eq:SDPN}), Model IV, is close to the ML performance. Similar
to the SDP relaxation model \eqref{eq:SDPF}, the algorithm based on
Model IV outperforms the SD algorithm in terms of the worst case
complexity (polynomial vs. exponential). Furthermore, by using the
LLL lattice basis reduction before the proposed SDP model, the
complexity is reduced, with an acceptable degradation in the
performance (as shown in Simulation Results section).

As a final note, we must emphasis that in the complexity analysis
for model \eqref{eq:SDPN}, we have considered all the non-negative
constraints. This suggests that the complexity of this model is not
tractable. However, it is not required to consider all the
non-negative constraints. In order to implement this model more
efficiently, we can solve the SDP relaxation \eqref{eq:SDPF}, and
solve the SDP relaxation \eqref{eq:SDPN} with only the \emph{most
violated constraints}. These constraints correspond to those
positions in matrix ${\bf Y}$ where their values are the minimum
negative numbers. Implementing Model IV based on the most violated
constraints reduces the complexity to almost a number of times more
complex compared to the Model III.

\section{Conclusion}

A method for quasi-maximum likelihood decoding based on two
semi-definite relaxation models is introduced. The proposed
semi-definite relaxation models provide a wealth of trade-off
between the complexity and the performance. The strongest model
provides a near-ML performance with polynomial-time worst-case
complexity (unlike the SD that has exponential-time complexity).
Moreover, the soft decoding method based on the proposed models is
investigated. By using lattice basis reduction the complexity of the
decoding methods based on theses models is reduced.


\section*{acknowledgement}

The authors would like to thank Prof. H. Wolkowicz for his
tremendous help and support throughout this work. Also, we would
like to thank Steve Benson and Futakata Yoshiaki in providing help
and support in using the DSDP and SDPA software.

\begin{appendices}


\section{Lagrangian Duality}\label{app:lag}

In this appendix, we show that Lagrangian duality can be used to
derive the SDP relaxation problem (\ref{eq:SemiR}). We first dualize
the constraints of (\ref{eq:semiprob}), and then derive the SDP
relaxation from the dual of the homogenized Lagrangian dual.
Finally, we project the obtained relaxation onto the minimal face.
The resulting relaxation is equivalent to the relaxation
(\ref{eq:SemiR}).

Consider the minimization problem in (\ref{eq:semiprob}). According
to \cite{WoSaVa00}, for an accurate semi-definite solution, zero-one
constraints should be formulated as quadratic constraints.
Therefore,
\begin{align}\label{eq:QAP0}
\nonumber \min \;\; &\mathbf{u}^T \mathbf{Q}
\mathbf{u} + 2 \mathbf{c}^T \mathbf{u} \\
\nonumber
s.t. \;\; &\| \mathbf{A}\mathbf{u}-\mathbf{e}_N\|^2=0 \\
&u_i^2=u_i \;\; \forall i=1,\cdots,n .
\end{align}
First, the constraints are added to the objective function using
lagrange multipliers $ \lambda $ and $ \mathbf{\tilde{w}} = \left[
\tilde{w}_1, \cdots, \tilde{w}_n \right]^T $:
\begin{align}
\nonumber \mu_{\mathcal{O}} = \min_{\mathbf{u}}
\max_{\lambda,\mathbf{\tilde{w}}} &\left\lbrace \mathbf{u}^T
\mathbf{Q} \mathbf{u} + 2 \mathbf{c}^T \mathbf{u}\right.\\
\nonumber &+ \lambda \left( \mathbf{u}^T \mathbf{A}^T \mathbf{A}
\mathbf{u} -
2\mathbf{e}_N^T \mathbf{A} \mathbf{u} + \mathbf{e}_N^T \mathbf{e}_N \right)\\
& + \sum_{i=1}^n \tilde{w}_i \left( u_i^2 - u_i \right)
\left.\right\rbrace.
\end{align}
Interchanging min and max yields
\begin{align}
\nonumber \mu_{\mathcal{O}} \geq \mu_{\mathcal{L}} =
\max_{\lambda,\mathbf{\tilde{w}}} \min_{\mathbf{ u}} &\left\lbrace
\mathbf{u}^T \mathbf{Q} \mathbf{u} + 2 \mathbf{c}^T \mathbf{u}\right.\\
\nonumber &+ \lambda \left( \mathbf{u}^T \mathbf{A}^T \mathbf{A}
\mathbf{ u} - 2\mathbf{e}_N^T \mathbf{A} \mathbf{u} + \mathbf{e}_N^T
\mathbf{e}_N \right)\\
& + \sum_{i=1}^n \tilde{w}_i \left( u_i^2 - u_i \right) \left.
\right\rbrace.
\end{align}
Next, we homogenize the objective function by multiplying t with a
constrained scalar $ u_0 $ and then increasing the dimension of the
problem by $ 1 $. Homogenization simplifies the transition to a
semi-definite programming problem. Therefore, we have
\begin{align}
\nonumber \mu_{\mathcal{O}} \geq \mu_{\mathcal{L}} =
\max_{\lambda,\mathbf{\tilde{w}}} \min_{\mathbf{ u},u_0^2=1}
&\left\lbrace \mathbf{u}^T \left[ \mathbf{Q} + \lambda \mathbf{A}^T
\mathbf{A} + \textmd{Diag}(\mathbf{\tilde{w}}) \right] \mathbf{u}
\right. \\ \nonumber& - \left( 2\lambda\mathbf{e}_N^T\mathbf{A} -
2\mathbf{c}^T + \mathbf{\tilde{w}}^T \right)u_0\mathbf{u}\\ &+
\lambda \mathbf{e}_N^T \mathbf{e}_N \left.\right\rbrace ,
\end{align}
where $ \textmd{Diag}(\mathbf{\tilde{w}}) $ is a diagonal matrix
with $ \mathbf{\tilde{w}} $ as its diagonal elements. Introducing a
Lagrange multiplier $ w_0 $ for the constraint on $ u_0 $, we obtain
the lower bound $ \mu_{\mathcal{R}} $
\begin{align}
\nonumber \mu_{\mathcal{O}} \geq \mu_{\mathcal{L}} \geq
\mu_{\mathcal{R}}\hspace{-4pt} = \hspace{-4pt}\max_{ \lambda,
\mathbf{ \tilde{w}}, w_0} &\min_{u_0,\mathbf{u}} \left\lbrace
\mathbf{ u}^T \hspace{-3pt} \left[ \mathbf{Q} + \lambda \mathbf{A}^T
\hspace{-2pt} \mathbf{A} + \textmd{Diag}(\mathbf{\tilde{w}}) \right]
\hspace{-3pt} \mathbf{u}
\right.\\
\nonumber & - \left( 2\lambda \mathbf{e}_N^T \mathbf{A} -
2\mathbf{c}^T + \mathbf{\tilde{w}}^T \right)u_0\mathbf{u}\\
& + \lambda \mathbf{e}_N^T \mathbf{e}_Nu_0^2 + w_0 \left( u_0^2 - 1
\right)\left.\right\rbrace.
\end{align}
Note that both inequalities can be strict, i.e. there can be duality
gaps in each of the Lagrangian relaxations. Also, the multiplication
of $ \lambda \mathbf{e}_N^T \mathbf{e}_N $ by $ u_0^2 $ is a
multiplication by $ 1 $. Now, by grouping the quadratic, linear, and
constant terms together and defining $ \mathbf{\tilde{u}}^T = \left[
u_0, \mathbf{u}^T \right]^T $ and $ \mathbf{w}^T = \left[ w_0,
\mathbf{\tilde{w}}^T \right]^T $, we obtain
\begin{equation}\label{eq:mur}
\mu_{\mathcal{R}} \hspace{-4pt} = \max_{\lambda,\mathbf{w}}
\min_{\tilde{\mathbf{u}}} \left\lbrace \tilde{\mathbf{u}}^T
\hspace{-2pt} \left[ \mathcal{L}_{\mathbf{Q}} + \textmd{Arrow}(
\mathbf{w}) + \lambda \mathcal{L}_{\lambda} \right] \hspace{-2pt}
\mathbf{u} - w_0 \right\rbrace,
\end{equation}
where
\begin{align}
\nonumber &\mathcal{L}_{\lambda} = \left[\hspace{-5pt}
\begin{tabular}{cc}
$ \mathbf{e}_N^T \mathbf{e}_N $ & \hspace{-5pt}$ -\mathbf{e}_N^T\mathbf{A} $\\
$ -\mathbf{A}^T\mathbf{e}_N $ & \hspace{-5pt}$ \mathbf{A}^T\mathbf{A} $\\
\end{tabular} \hspace{-5pt}
\right] = \left[\hspace{-5pt}
\begin{tabular}{cc}
$ N $ & \hspace{-5pt}$ -\mathbf{e}^T_K \otimes \mathbf{e}^T_N $\\
$ -\mathbf{e}_K \otimes \mathbf{e}_N $ & \hspace{-5pt}$ \mathbf{I}_N \otimes (\mathbf{e}_K\mathbf{e}^T_K) $ \\
\end{tabular}\hspace{-5pt}
\right],\\
\nonumber&\textmd{Arrow}(\mathbf{w})=\left[
\begin{tabular}{cc}
$ w_0 $ & $ -\frac{1}{2}\mathbf{w}^T_{1:n} $\\
$ -\frac{1}{2}\mathbf{w}_{1:n} $ & $ \textmd{Diag}(\mathbf{w}_{1:n}) $\\
\end{tabular}
\right],\\
&\textmd{and }\mathcal{L}_{\mathbf{Q}}=\left[
\begin{tabular}{cc}
$ 0 $ & $ \mathbf{c}^T $\\
$ \mathbf{c} $ & $ \mathbf{Q} $\\
\end{tabular}
\right].
\end{align}

Note that we will refer to the additional row and column generated
by the homogenization of the problem as the 0-th row and column.
There is a hidden semi-definite constraint in (\ref{eq:mur}), i.e.
the inner minimization problem is bounded below only if the Hessian
of the quadratic form is positive semi-definite. In this case, the
quadratic form has minimum value $ 0 $. This yields the following
SDP problem:
\begin{align}\label{eq:D0}
\nonumber\max \;\; &- w_0 \\
s.t. \;\; &\mathcal{L}_{\mathbf{Q}}+ \textmd{Arrow}(\mathbf{w}) +
\lambda \mathcal{L}_{ \lambda} \succeq 0.
\end{align}
We now obtain our desired SDP relaxation of (\ref{eq:QAP0}) as the
Lagrangian dual of (\ref{eq:D0}). By Introducing the $ (n+1) \times
(n+1) $ dual matrix variable $ \mathbf{Y} \succeq 0 $, the dual
program to the SDP (\ref{eq:D0}) would be
\begin{align}\label{eq:SDP0}
\nonumber \min \;\; &\textmd{trace } \mathcal{L}_{\mathbf{Q}} \mathbf{Y}\\
\nonumber s.t. \;\; &{\rm diag}(\mathbf{Y})= (1, \mathbf{Y}_{0,1:n})^T\\
\nonumber \;\; &\textmd{trace} \mathbf{Y}\mathcal{L}_{\lambda} = 0\\
\;\; &\mathbf{Y} \succeq 0,
\end{align}
where the first constraint represents the zero-one constraints in
(\ref{eq:QAP0}) by guaranteeing that the diagonal and 0-th column
(row) are identical (matrix $ {\bf Y} $ is indexed from 0); and the
constraint ${\bf Au}={\bf e}_N$ is represented by the constraint $
\textmd{trace} \mathbf{Y}\mathcal{L}_{\lambda} = 0 $. Note that if
the matrix $ \mathbf{Y} $ is restricted to be rank-one in
(\ref{eq:SDP0}), i.e.
\begin{displaymath}
\mathbf{Y}=\left[
\begin{tabular}{c}
$ 1 $\\
$ \mathbf{u} $\\
\end{tabular}
\right] \left[1 \; \mathbf{u}^T \right],
\end{displaymath}
for some $ \mathbf{u} \in \mathbb{R}^n $, then the optimal solution
of (\ref{eq:SDP0}) provides the optimal solution, $ \mathbf{u} $,
for (\ref{eq:QAP0}).

Since the matrix $ \mathcal{L}_{\lambda} \neq 0 $ is a positive
semi-definite matrix; therefore, to satisfy the constraint in
(\ref{eq:SDP0}), $ \mathbf{Y} $ has to be singular. This means the
feasible set of the primal problem in (\ref{eq:SDP0}) has no
interior \cite{ZhKaReWo96} and an interior-point method may never
converge. However, a simple structured matrix can be found in the
relative interior of the feasible set in order to project (and
regularize) the problem into a smaller dimension.

As mentioned before, the rank-one matrices are the extreme points of
the feasible set of the problem in (\ref{eq:SDP0}) and the minimal
face of the feasible set that contains all these points shall be
found \cite{ZhKaReWo96}.

From Theorems \ref{lem:SetP} and \ref{thm:bary}, we conclude that $
\mathbf{Y} \succeq 0 $ is in the minimal face if and only if $
\mathbf{Y} = \mathbf{\hat{V}R\hat{V}}^T $, for some $ \mathbf{R}
\succeq 0 $. By substituting $ \mathbf{\hat{V}R\hat{V}}^T $ for $
\mathbf{Y} $ in the SDP relaxation (\ref{eq:SDP0}), we get the
following projected SDP relaxation which is the same as the SDP
relaxation in (\ref{eq:SemiR}):
\begin{align}\label{eq:QAPR1}
\nonumber
\mu_{R1} = \min \;\; &\textmd{trace } (\mathbf{\hat{V}}^T\mathcal{L}_{\mathbf{Q}} \mathbf {\hat{V}}) \mathbf{R}\\
\nonumber \textmd{s.t. }\;
&\textmd{diag}(\mathbf{\hat{V}R\hat{V}}^T)= (1,
(\mathbf{\hat{V}R\hat{V}}^T)_{0,1:n})^T\\ & \mathbf{R} \succeq 0.
\end{align}
Note that the constraint $ \textmd{trace} (\mathbf{\hat{V}}^T
\mathcal{L}_{\lambda} \mathbf{\hat{V}}) \mathbf{R} = 0 $ can be
dropped since it is always satisfied, i.e. $ \mathcal{L}_{\lambda}
\mathbf{\hat{V}} =0 $.

\section{Proofs}\label{app:proof}

\subsection{Lemma \ref{new}}

Let $ {\bf X} \in {\mathcal M}_{K\times N}$ and $ {\bf e}^T_K {\bf
X} = {\bf e}^T_N$. Since ${\bf V}_{K \times (K-1)} $ is a $K\times
(K-1)$ matrix containing a basis of the orthogonal complement of the
vector of all ones, i.e., $ {\bf V}_{K \times (K-1)}^T {\bf
e}_K={\bf 0}$, and
\begin{equation}
{\bf e}^T_K {\bf F}_{K\times N} = {\bf e}^T_N,
\end{equation}
we have
\begin{equation}
{\bf X} = {\bf F}_{K \times N} + {\bf V}_{K \times (K-1)}{\bf Z},
\end{equation}
where ${\bf Z}\in {\mathcal M}_{(K-1)\times N}$. From
{\setlength\arraycolsep{2pt}
\begin{align}
\nonumber {\bf F}_{K\times N} &= \frac{1}{K}( {\bf E}_{K\times N} -
{\bf V}_{K \times (K-1)} {\bf E}_{(K-1)\times N})\\
& = \left [
\begin{array}{c}
{\bf 0}_{(K-1)\times N} \\ \hline {\bf e}_{N}^{T}
\end{array} \right ],
\end{align}}
and
\begin{equation}
{\bf V}_{K \times (K-1)} {\bf Z} = \left [
\begin{array}{c}
{\bf Z} \\ \hline -{\bf e}^T_{K-1} {\bf Z}
\end{array}
\right ]
\end{equation}
it follows that ${\bf Z} = {\bf X}(1:(K-1), 1:N)$.

\subsection{Theorem \ref{lem:SetP}}

Let ${\bf Y}\in {\mathcal F}$ be an extreme point of ${\mathcal F}$,
i.e.
\begin{equation}\label{eq:matY}
{\bf Y} = {\bf Y}_{\bf u} = \left [ \begin{array}{c|c} 1 & {\bf x}^{T} \\
\hline {\bf x} & {\bf xx}^{T}
\end{array} \right ],
\end{equation}
for some  ${\bf x}=\kvec({\bf X})$, ${\bf X} \in {\mathcal E}_{K
\times N}$. From Lemma \ref{new}, it follows that every matrix ${\bf
X} \in {\mathcal E}_{K \times N}$ is of the form ${\bf X}= {\bf
F}_{K \times N} + {\bf V}_{K \times (K-1)}\tilde {\bf X}$ where
$\tilde {\bf X} = {\bf X}(1:K-1,1:N)$. From the properties of the
Kronecker product (see \cite{Gra81}), it follows
\begin{align}
\nonumber {\bf x} = \kvec({\bf X})&=\frac{1}{K} ({\bf e}_{KN} -
({\bf I}_N \otimes {\bf V}_{K \times (K-1)} ){\bf e}_{(K-1)N} )\\ &+
({\bf I}_N \otimes {\bf V}_{K \times (K-1)})\tilde {\bf x},
\end{align}
where $\tilde {\bf x}=\kvec(\tilde {\bf X})$. Let ${\bf p}^T :=
\left [
\begin{array}{cc} 1 & \tilde {\bf x}^T \end{array} \right ]$ and
\begin{equation} \label{W:def}
{\bf W} := \left[  \frac{1}{K} ( {\bf e}_{KN} - ({\bf I}_N \otimes
{\bf V}_{K \times (K-1)}){\bf e}_{(K-1)\times N}), {\bf I}_N \otimes
{\bf V}_{K \times (K-1)} \right ].
\end{equation}
Therefore, $ {\bf x}={\bf Wp}$, and
\begin{equation}
{\bf Y}= \left [ \begin{array}{c|c}
1  & {\bf p}^T{\bf W}^{T} \\[1.ex]\hline
{\bf Wp} & {\bf W pp}^{T}{\bf W}^{T}
\end{array} \right ] =
\hat {\bf V} {\bf R}\hat {\bf V}^{T},
\end{equation}
where ${\bf R}:={\bf pp}^{T}$, i.e.
\begin{equation} \label{def:matR}
{\bf R} = \left [
\begin{array}{c|c}
1 & \tilde {\bf x}^{T} \\
\hline \tilde {\bf x}  & \tilde {\bf x} \tilde {\bf x}^{T}
\end{array} \right ] \succeq {\bf 0}.
\end{equation}
Since $\tilde {\bf x}$ is a binary vector, it follows that $r_{ij}
\in \{0,1 \},~\forall i,j \in \{0,\ldots, N(K-1) \}$, and ${\rm
diag}(\tilde {\bf x} \tilde {\bf x}^{T})=\tilde {\bf x}$. The proof
follows analogously  for any convex combination of the extreme
points from ${\mathcal F}$.

\subsection{Theorem \ref{thm:bary}}

Fix $ \mathbf{u} \in \mathcal{U} $ and let

\begin{center}
\begin{math}
\mathbf{Y} = {\bf Y}_{\bf u} = \left[
\begin{tabular}{c}
$ 1 $\\
$ \mathbf{u} $
\end{tabular}
\right] \left[ 1 \; \mathbf{u}^T \right].
\end{math}
\end{center}
Considering the constraint on this vector in (\ref{eq:semisum}), we
divide ${\bf u}$ into $ N $ \textit{sub-vectors} of length $ K $. In
each sub-vector all the elements are zero except one of the elements
which is one. Therefore, there are $ K^N $ different binary vectors
in the set $ \mathcal{U} $. Consider the entries of the 0-th row of
$ \mathbf{Y} $. Note that $ y_{0,j} = 1 $ means that the $ j $-th
element of $ \mathbf{u} $ is 1. In addition, there is only one
element equal to $ 1 $ in each sub-vector. Therefore, there are $
K^{N-1} $ such vectors, and the components of the 0-th row of $
\hat{\mathbf{Y}} $ are given by
\begin{center}
\begin{math}
\hat{y}_{0,j} = \dfrac{1}{K^N}K^{N-1} = \dfrac{1}{K}.
\end{math}
\end{center}
Now consider the entries of $ \mathbf{Y} $ in the other rows, $
Y_{i,j} $.
\begin{enumerate}
\item If $ i=j $, then, $ y_{i,j} =1 $ means that the $ i $-th
element of the vector $ \mathbf{u} $ is 1 and there are $ K^{N-1} $
such vectors; therefore, the diagonal elements are
\begin{center}
\begin{math}
\hat{y}_{i,i} = \dfrac{1}{K^N}K^{N-1} = \dfrac{1}{K}.
\end{math}
\end{center}
\item If $ i=K(p-1)+q, \; j=K(p-1) + r, q \neq r, q,r \in \lbrace
1,\cdots,K \rbrace , p \in \lbrace 1, \cdots, N\rbrace $, i.e. the
element is an off-diagonal element in a diagonal block, then, $
y_{i,j} =1 $ means that the $ i $-th and the $ j $-th elements of $
\mathbf{u} $ in a sub-vector should be 1 and this is not possible.
Therefore, this element is always zero.
\item Otherwise, we
consider the elements of the off-diagonal blocks of $ \mathbf{Y} $.
Then, $ y_{i,j} =1 $ means that the $ i $-th and the $ j $-th
elements of $ \mathbf{u} $ in two different sub-vectors are 1 and
there are $ K^{N-2} $ such vectors; therefore, the elements of the
off-diagonal blocks are
\begin{center}
\begin{math}
\hat{y}_{i,i} = \dfrac{1}{K^N}K^{N-2} = \dfrac{1}{K^2}.
\end{math}
\end{center}
\end{enumerate}

\noindent This proves the representation of $ \mathbf{\hat{Y}} $ in
(i) in Theorem \ref{thm:bary}.

It can be easily shown that
\begin{equation}\label{eq:eigen}
\left[\hspace{-2pt}
\begin{tabular}{c|c}
$ 1 $ & $ {\bf 0}_n^T $\\
\hline $ -\frac{1}{K} \mathbf{e}_n $ & $ \mathbf{I}_n $
\end{tabular}\hspace{-2pt}\right] \mathbf{\hat{Y}} \left[\hspace{-2pt}
\begin{tabular}{c|c}
$ 1 $ & $ -\frac{1}{K} \mathbf{e}^T_n $\\
\hline $ {\bf 0}_n $ & $ \mathbf{I}_n $
\end{tabular}\hspace{-2pt}\right]=\left[\hspace{-2pt}
\begin{tabular}{c|c}
$ 1 $ & $ {\bf 0}_n^T $\\
\hline $ {\bf 0}_n $ & $ \hat{\mathbf{W}} $
\end{tabular}\hspace{-2pt}\right],
\end{equation}
where $ \hat{\mathbf{W}} = \dfrac{1}{K^2} \mathbf{I}_N \otimes
\left( K\mathbf{I}_K - \mathbf{E}_K \right) $. Note that $
\textmd{rank}(\mathbf{\hat{Y}})  = 1 +
\textmd{rank}(\mathbf{\hat{W}}) $.\\
The eigenvalues of $ \mathbf{I}_N $ are 1, with multiplicity $ N $
and the eigenvalues of $ K\mathbf{I}_K - \mathbf{E}_K $ are $ K $,
with multiplicity $ K-1 $, and 0. Note that the eigenvalues of a
Kronecker product are given by the Kronecker product of the
eigenvalues \cite{Gra81}. Therefore, the eigenvalues of $
\hat{\mathbf{W}} $ are $ \frac{1}{K} $, with multiplicity $ N(K-1)
$, and 0, with multiplicity $ N $. Therefore, we have
\begin{equation}\label{eq:rank}
\textmd{rank}(\mathbf{\hat{Y}})  = 1 + \textmd{rank}
(\mathbf{\hat{W}}) = N(K-1) + 1.
\end{equation}
This proves (ii) in Theorem \ref{thm:bary}.

By (\ref{eq:eigen}) and (\ref{eq:rank}), we can easily see that the
eigenvalues of $ \mathbf{\hat{Y}} $ are $ \dfrac{1}{K} $, with
multiplicity $ N(K-1) $, $ \dfrac{K+N}{K} $, and 0, with
multiplicity $ N $. This proves (iii) in Theorem \ref{thm:bary}.

The only constraint that defines the minimal face is ${\bf Au}={\bf
b}$. By multiplying of both sides by ${\bf u}^T$ and using the fact
that ${\bf u}$ is a binary vector, we obtain
\begin{equation}
{\bf Tuu}^T = {\bf e}_N\left({\rm diag}({\bf uu}^T)\right)^T.
\end{equation}
This condition is equivalent to
\begin{equation}
{\bf T}\hat{\bf Y}={\bf 0}.
\end{equation}
Note that ${\rm rank}({\bf T})=N$. Therefore, we have $$
\mathcal{N}(\hat{\bf Y})=\lbrace {\bf u}: {\bf u} \in
\mathcal{R}({\bf T}^T) \rbrace .$$ This proves (iv) in Theorem
\ref{thm:bary}.

Since $ \textmd{rank}(\mathbf{\hat{V}}) = N(K-1)+1 $ and using
Theorem \ref{lem:SetP}, the columns of $ \mathbf{\hat{V}} $ span the
range space of $ \mathbf{\hat{Y}} $. This proves (v) in Theorem
\ref{thm:bary}.

\section{Randomization Algorithm}\label{app:rand}

In this appendix, we present two randomization algorithms to
transform ${\bf Y}$ to a binary rank-one matrix.

\subsection{Algorithm I}

Goemans and Williamson \cite{GoeWil95} introduced an algorithm that
randomly transforms an SDP relaxation solution to a rank-one
solution. This approach is used in \cite{StLuWo03} for the quasi
maximum likelihood decoding of a PSK signalling. This technique is
based on expressing the BPSK symbols by $\{-1,1\}$ elements. After
solving the relaxation problem in \cite{StLuWo03}, the Cholesky
factorization is applied to the $n\times n$ matrix $\bf Y$ and the
Cholesky factor ${\bf V}=[{\bf v}_1,\ldots,{\bf v}_n]$ is computed,
i.e. ${\bf Y}={\bf VV}^T$. In \cite{StLuWo03}, it is observed that
one can approximate the solution of the distance minimization
problem, $\bf u$, using $\bf V$, i.e. $u_i$ is approximated using
${\bf v}_i$. Thus, the assignment of $-1$ or $1$ to the vectors
$\{{\bf v}_1,\ldots, {\bf v}_n \}$ is equivalent to specifying the
elements of $\bf u$.
\begin{figure}[tbhp]
\centering \bf
\includegraphics[scale=0.7]{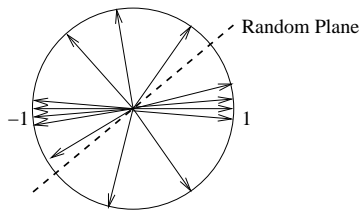}
\caption{Representation of the randomization algorithm in
\cite{GoeWil95}}\label{fig:rand}
\end{figure}

It is shown that norms of the vectors $\{{\bf v}_1,\ldots, {\bf v}_n
\}$ are one, and they are inside an $n$--dimensional unit sphere
\cite{StLuWo03}, see Fig. \ref{fig:rand}. These vectors should be
classified in two different groups corresponding to $1$ and $-1$. In
order to assign $-1$ or $1$ to these vectors, the randomization
procedure generates a random vector uniformly distributed in the
sphere. This vector defines a plane crossing the origin. Among given
vectors ${\bf v}_i$, $i=1,\ldots n$, all the vectors at one side of
the plane are assigned to $1$ and the rest are assigned to $-1$, as
shown in Fig. \ref{fig:rand}. This procedure is repeated several
times and the vector $\bf u$ resulting in the lowest objective
function is selected as the answer.

In our proposed approach, the variables are binary numbers. In order
to implement the randomization procedure of \cite{GoeWil95}, we
bijectively map the computed solution of the $\{0,1\}$ SDP
formulation to the solution of the corresponding $\{-1,1\}$ SDP
formulation. More precisely, we use the following mapping:
\begin{eqnarray}\label{eq:tran}
\nonumber {\bf M}=\left[
\begin{array}{c|ccc}
1&0&\cdots&0\\
\hline
-1&2&\cdots&0\\
\vdots&\vdots&\ddots&\vdots\\
-1&0&\cdots&2
\end{array}\right],\\
{\bf Y}_{\{-1,1\}} = {\bf MY}_{\{0,1\}}{\bf M}^T,
\end{eqnarray}
where ${\bf Y}_{\{0,1\}}$ is the resulting matrix from the
relaxation model (\ref{eq:SDPF}) or (\ref{eq:SDPN}) and ${\bf
Y}_{\{-1,1\}}$ is its corresponding matrix with $\{-1,1\}$ elements.
Using (\ref{eq:tran}), the solution for (\ref{eq:semiprob}) can be
computed using a similar randomization method as in \cite{StLuWo03}.
The computational complexity of this randomization algorithm is
polynomial \cite{StLuWo03}.

Considering zero-one elements in our problem, we propose a new
randomization procedure inspired by \cite{GoeWil95}. This algorithm
can be applied to $\{0,1\}$ formulation directly. Therefore, the
complexity of the whole randomization procedure is reduced, since
the preprocessing step, i.e. bijective mapping in (\ref{eq:tran}),
is omitted.

\subsection{Algorithm II}
\begin{figure}[tbhp]
\centering \bf
\includegraphics[scale=0.85]{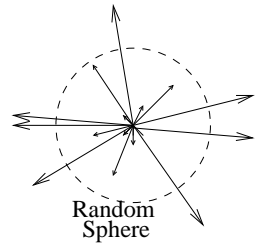}
\caption{Graphic representation for the proposed randomization
algorithm}\label{fig:rand01}
\end{figure}

After solving the relaxation model (\ref{eq:SDPF}) or
(\ref{eq:SDPN}), the Cholesky factorization of $\bf Y$ results in a
matrix $\bf V=[{\bf v}_1,\ldots,{\bf v}_n]$ such that ${\bf Y}={\bf
VV}^T$. The matrix ${\bf Y}$ is neither binary nor rank-one.
Therefore, norms of the resulting vectors ${\bf v}_i$ are between
zero and one. These vectors are depicted in Fig. \ref{fig:rand01}.
Intuitively, a sphere with a random radius uniformly distributed
between zero and one has the same functionality as the random plane
in Fig. \ref{fig:rand}.

In order to assign $0$ or $1$ to these vectors, the randomization
procedure generates a random number, uniformly distributed between 0
and 1, as the radius of the sphere. Among given vectors ${\bf v}_i$,
$i=1,\ldots, n$, all the vectors whose norms are larger than this
number are assigned to $1$ and the rest are assigned to $0$. In
another variation of this algorithm, the radius of the sphere can be
fixed, and norms of these vectors are multiplied by a random number.
This procedure is repeated several times and the vector $\bf u$
resulting in the smallest objective function value in
(\ref{eq:semiprob}) is selected as the solution. Simulation results
confirm that the proposed method results in a slightly better
performance for the lattice decoding problem compared to the first
algorithm.
Also, the computational complexity of the randomization algorithm is
decreased, due to the removal of the preprocessing step in
(\ref{eq:tran}). It is worth mentioning that, according to
(\ref{eq:matY}) and (\ref{def:matR}), the randomization procedure
can be implemented for the matrix ${\bf R}$, which results in
further reduction in the computational complexity.


\end{appendices}

\end{document}